\documentclass[prb,twocolumn,showpacs,superscriptaddress,floatfix,amsmath,amssymb]{revtex4}

\usepackage{graphicx} 
\usepackage{color}
\usepackage{dcolumn} 
\usepackage{bm} 

\newcommand{\be}{\begin{equation}}
\newcommand{\ee}{\end{equation}}
\newcommand{\bea}{\begin{eqnarray}}
\newcommand{\eea}{\end{eqnarray}}
\newcommand{\reff}{\text{eff}}
\newcommand{\up}{\uparrow}
\newcommand{\down}{\downarrow}

\newcommand{\verysmallfig}{0.4\textwidth}
\newcommand{\smallfig}{0.45\textwidth}

\begin{document}

\title{Nature of the metal-insulator transition in the 
half--filled $t-t^{\prime}$ Hubbard chain}

\date{June 28, 2006}

\author{G.\ I.\ Japaridze}
\affiliation{Max-Planck-Institut f\"ur  Physik komplexer Systeme,
N\"otnitzer Str.\ 38, D-01187 Dresden, Germany}
\altaffiliation{Permanent address:
 Andronikashvili Institute  of  Physics, Georgian Academy of Sciences,
 Tamarashvili 6, 0177 Tbilisi, Georgia}
\author{R.\ M.\ Noack}
\affiliation{Fachbereich Physik, Philipps-Universit\"at Marburg, 
D-35032 Marburg, Germany} 
\author{D.\ Baeriswyl}  
\affiliation{D\'epartement de physique, Universit\'{e} de Fribourg, 
CH-1700 Fribourg, Switzerland}

\begin{abstract}

We study the quantum phase transition from an insulator to a metal 
realized at $t^{\prime}=t^{\prime}_{c}  > 0.5t$ in the ground state of the
half-filled Hubbard chain with both nearest-neighbor ($t$) and
next-nearest-neighbor ($t^{\prime}$) hopping. 
The study is carried out
using the bosonization approach and density matrix renormalization
group calculations. 
An effective low-energy Hamiltonian that describes the
insulator-metal transition is derived.
We find that the gross features of
the phase diagram are well-described by the standard theory of
commensurate-incommensurate transitions in a wide range of parameters.
We also obtain an analytical expression for the insulator-metal
transition line $t^{\prime}_{c}(U,t)$. 
We argue that close to the 
insulator-metal transition line, a crossover to a regime
corresponding to an infinite-order transition takes place.
We present results of density-matrix-renormalization-group
calculations of spin and charge distribution in various sectors of
the phase diagram.
The numerical results support the picture derived from the effective
theory and give evidence for the complete separation of the transitions
involving spin and charge degrees of freedom.

\end{abstract}

\maketitle

\section{Introduction}

During the last decades, the Mott metal-insulator transition has
been the subject of great interest.
\cite{Mott_Book_90,Gebhard_Book_97, IFT_98}
In the canonical model for this transition -- the single-band Hubbard
model --  the origin of the insulating behavior is the on-site Coulomb
repulsion between electrons. 
For an average density of one electron per site, the transition from
the metallic to the insulating phase is expected to occur when the
electron-electron interaction strength $U$ is of the order of the
delocalization energy (which is a few times the hopping amplitude
$t$). 
The critical value $(U/t)_c$ turns out to be quite independent of the
specific band structure.\cite{DBM_97}
It is important to recall that the Mott transition is often preceded
by antiferromagnetic ordering, which usually leads to insulating
behavior and thus masks the Mott phenomenon.

While the underlying mechanism driving the Mott transition is by now
well understood, many questions remain open, especially about the
region close to the transition point where perturbative approaches
fail to provide reliable answers. The situation is more fortunate in
one dimension, where non-perturbative analytical methods together with
well-controlled numerical approaches allow in many cases to determine
both the ground state and the low-lying excited states.
\cite{LiebWu_68,AAO_1969,EFGKK_2005}  
However, even in one
dimension, apart from the exactly solvable cases, a full treatment of
the fundamental issues related to the Mott transition still
constitutes a hard and long-standing problem.

In this paper, we study the $t-t'$ Hubbard chain which includes both
nearest ($t$) and next-nearest-neighbor ($t'$) hopping terms. We limit
ourselves to an average density of one electron per site (the
half-filled band case). Depending on the ratio between $t'$ and $t$,
the system has two or four Fermi points. Correspondingly, it shows
one- or two-band behavior and has a rich phase diagram. Therefore, it
is not surprising that the model has been the subject of intensive
analytical and numerical studies, including a weak-coupling
renormalization group analysis,\cite{Fabrizio_96} DMRG
calculations for charge and spin gaps
\cite{Kuroki_97,Fabrizio_98,DaulNoack_98,DaulNoack_00,Torio_03},
the electric susceptibility,\cite{AebBaerNoack_01} the
momentum distribution function,\cite{Gros_01,Gros_02,Gros_04}
and the conductivity \cite{Torio_03} as well as, very
recently, a variational technique.\cite{Fabrizio_04}

Unfortunately, conflicting results have been reported for the
transition region, in particular regarding the character of the
transition, the number of different phases and the number of
gapless modes. 
In this paper, we hope to settle some of the unresolved
issues using a combined analytical and numerical analysis.  
We focus our attention on the metal-insulator transition as a function
of $t^{\prime}$ for a fixed on-site repulsion $U$. 
An effective continuum theory allows us to show that in the parameter
range $0.5t < t^{\prime}<t_{c}^{\prime}$ the system exhibits the
characteristic behavior of a commensurate-incommensurate transition.
\cite{C_IC_transition}
Close to the transition point, additional scattering processes
characteristic of two-band behavior set in. \cite{Fabrizio_96}
We argue that these processes induce a crossover to
Kosterlitz-Thouless type critical behavior, as found in 
Ref.\ \onlinecite{AebBaerNoack_01}.
Additional support for this conclusion is
provided by DMRG calculations for chains of up to $L=128$ sites  
and large $U$. 
The numerical analysis also allows us to study the 
gaps in the excitation spectrum as well as the charge and spin density
distributions.

The paper is organized as follows. In Section II, general properties of
the model including the strong coupling limit and the structure of the
phase diagram are briefly reviewed.  Section III shows that the
weak-coupling bosonization approach leads to the quantum sine-Gordon
field theory, the standard model for commensurate-incommensurate
transitions. In Section IV, the results of the numerical analysis are
presented, including the excitation spectrum and density
distributions.  Special attention is given to the regime of large $t'$
where the model represents a system of two coupled chains. The results
are summarized in Section V.

\section{\bf The $t-t'$ Hubbard chain}

The one-dimensional $t-t^{\prime}$ Hubbard model is defined by the
Hamiltonian 
\begin{eqnarray}\label{t1t2Umodel}
{\cal H}& = &
-t \,
\sum_{j,\sigma}
\left (c^{\dagger}_{j, \sigma}c^{\phantom{\dagger}}_{j+1,\sigma} \, + \,
c^{\dagger}_{j+1,\sigma}c^{\phantom{\dagger}}_{j, \sigma}\right )\nonumber \\
& + & t^{\prime} \, \sum_{j,\sigma} \left (c^{\dagger}_{j,
  \sigma}c^{\phantom{\dagger}}_{j+2,\sigma} \, + \,
c^{\dagger}_{j+2,\sigma}c^{\phantom{\dagger}}_{j, \sigma}\right ) \nonumber \\
& + & U \, \sum_{j,\sigma} \, (n_{j,\up}-1/2)\, (n_{j,\down}-1/2)\, ,
\end{eqnarray}
where $c^{\dagger}_{j, \sigma}$ $(c^{\phantom{\dagger}}_{j,\sigma})$
are electron creation (annihilation) operators on site $j$ and, with
spin projection $\sigma=\up,\down$,  
$n_{j,\sigma}= c^{\dagger}_{j,\sigma}c^{\phantom{\dagger}}_{j,\sigma}$,
and $U$ is the on-site Coulomb repulsion. 
\begin{figure}[tbh]
\includegraphics[width=\smallfig]{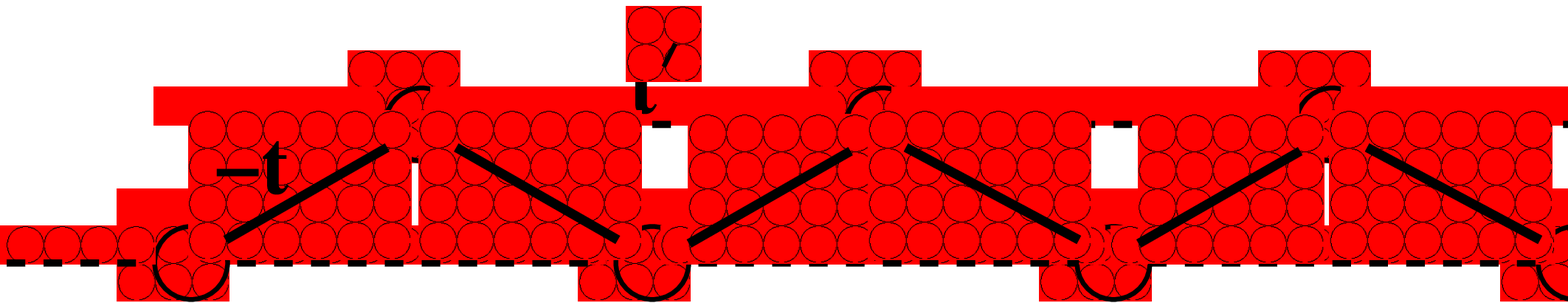}
\caption[DIAGONAL]{The $t-t^{\prime}$ Hubbard chain.}
\label{Fig:Fig1-lattice.eps}
\end{figure}

The model can be viewed either as a single chain with
both nearest- and next-nearest-neighbor hopping or, as illustrated in
Fig.\ref{Fig:Fig1-lattice.eps}, as a system of two coupled chains. 
The former view is appropriate for $t\gg t'$, the latter for $t'\gg t$.  

For $t^{\prime}=0$, we recover the ordinary Hubbard
model which is exactly (Bethe Ansatz) solvable.
\cite{LiebWu_68}
In the case of a half-filled band, the ground state is insulating
for arbitrary positive values of $U$; 
the charge excitation spectrum is gapped while the spin excitation
spectrum is gapless.\cite{LiebWu_68,AAO_1969}
For $U\ll t$ the charge gap $\Delta_c$ is exponentially small,
$\Delta_{c}  \approx \sqrt{U t}e^{-2\pi t/U}$, 
while  $\Delta_c\approx U$ for $U \gg t$.\cite{AAO_1969}

\begin{figure}[tbh]
\includegraphics[width=\verysmallfig]{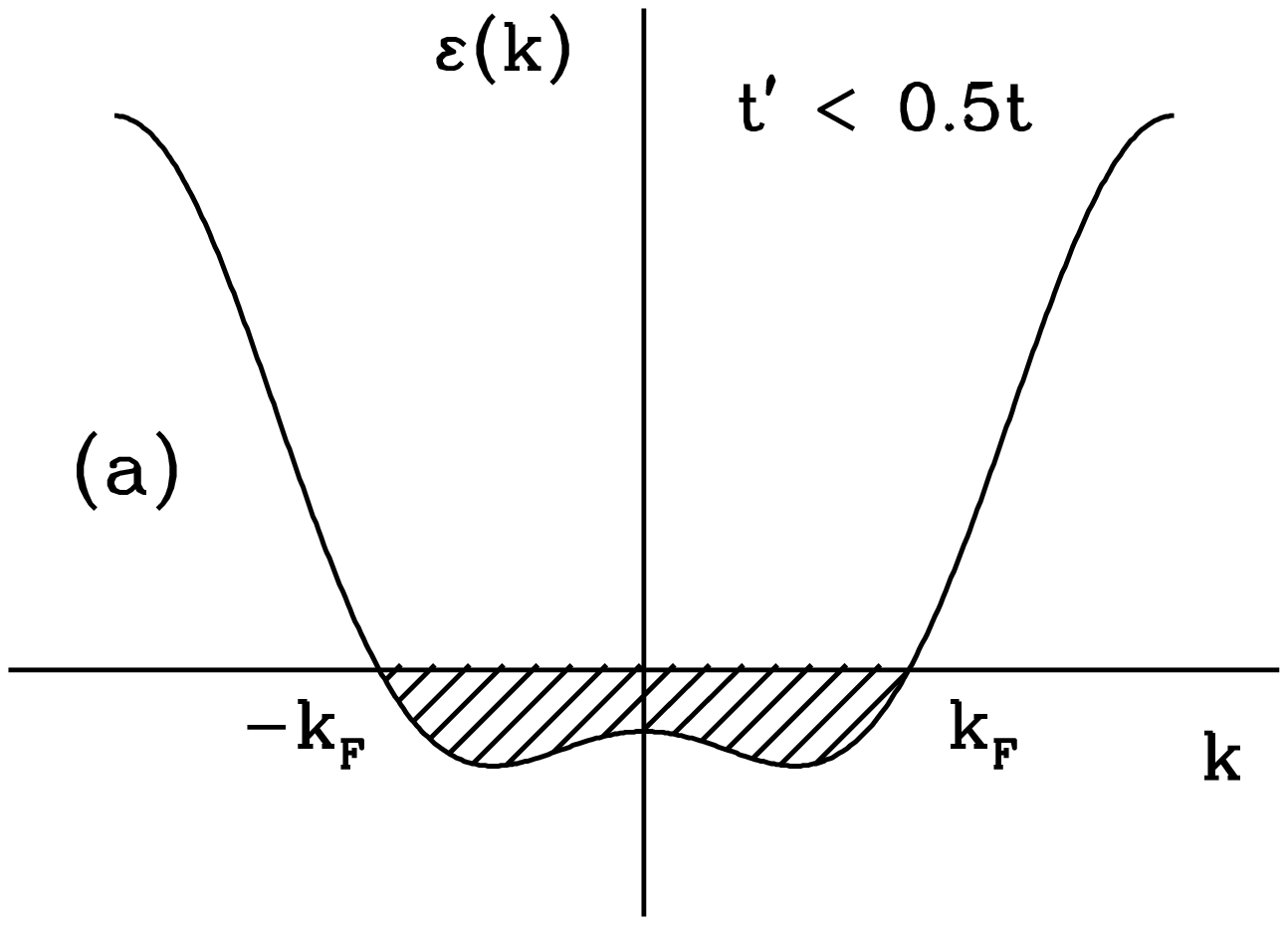}
\includegraphics[width=\verysmallfig]{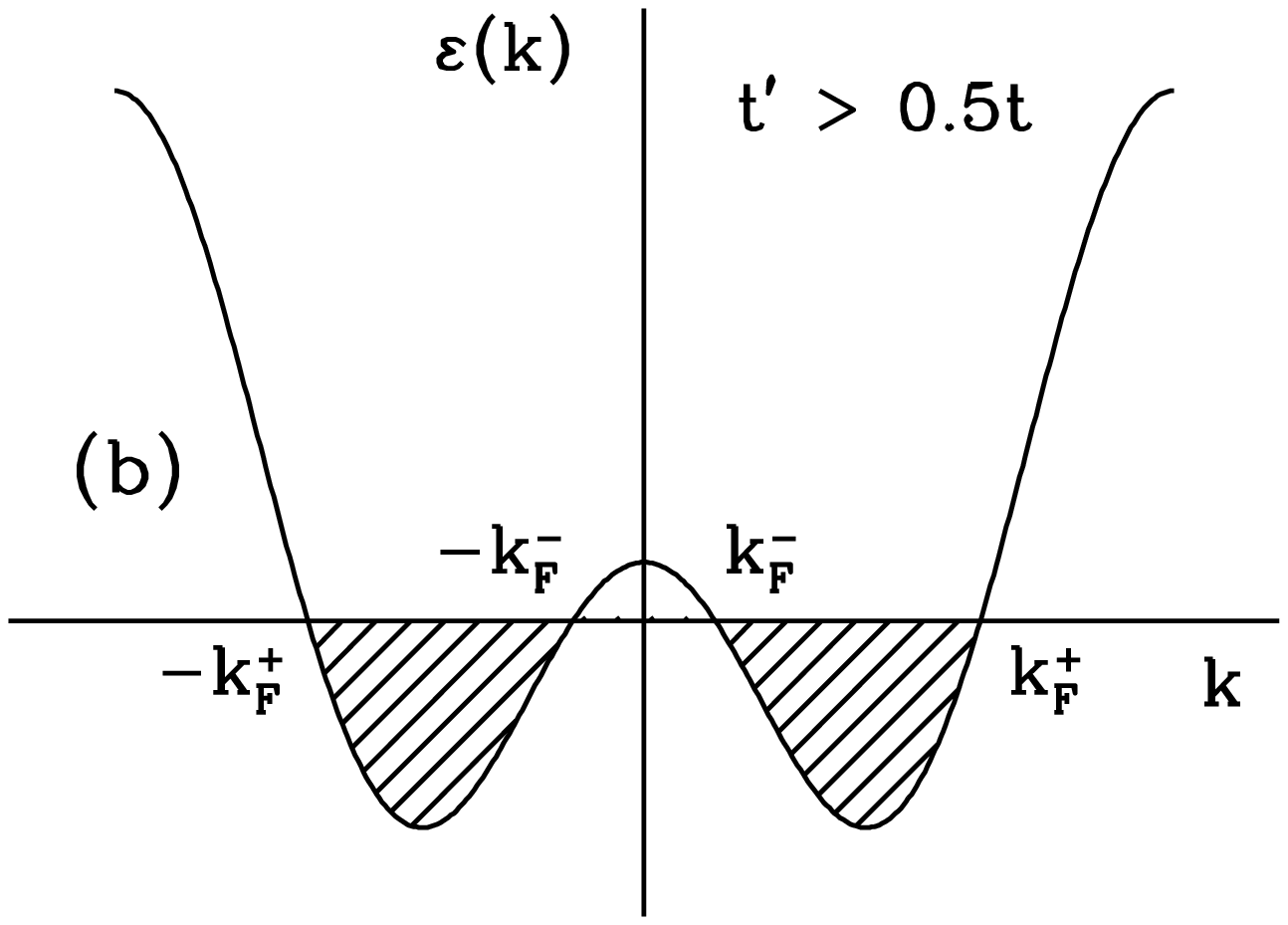}
\caption[DIAGONAL]{Single-particle dispersion relation of the
  $t-t^{\prime}$ chain for 
  (a) $t^{\prime}=0.4 t$ and  (b) $t^{\prime} = t$.}
\label{Fig:Fig2-bare-spectrum.eps}
\end{figure}

For $t^{\prime}\neq 0$ the model is no longer integrable except in the
non-interacting limit, $U=0$, where  $H$ is diagonalized by Fourier
transformation and has a single-electron spectrum 
\begin{equation}
\varepsilon(k) = -2t\cos k + 2t^{\prime}\cos 2k \, .
\end{equation}
For $t^{\prime}< 0.5t$, the electron band has two Fermi points at
$ k_{F}=\pm \pi/2$, separated from each other by the Umklapp vector
$q=\pi$ (see Fig.\ref{Fig:Fig2-bare-spectrum.eps}). 
In this case, a weak-coupling renormalization
group analysis \cite{Fabrizio_96} predicts the same behavior as
for $t^{\prime}=0$ because the 
Umklapp term of order $U$ is not modified; it again leads to the
dynamical generation  
of a charge gap for $U>0$, while the magnetic excitation spectrum
remains gapless.  

In the strong-coupling limit, $U\gg t,t^{\prime}$, the charge sector is
gapped, while the spin sector can be mapped onto a
frustrated Heisenberg chain
\be
  H = \sum_j \, \left( J \; {\bf S}_{j} \cdot {\bf S}_{j+1}
  \, + \, J^{\prime} \; {\bf S}_{j}\cdot {\bf S}_{j+2}  \right)
\label{FH-chain}
\ee
with $J=4t^{2}/U$ and $J^{\prime}=4{t^{\prime}}^{2}/U $. This model has been
extensively studied using a number of different analytical methods
\cite{Haldane_82,Eggert_96,WhiteAffleck_96} and has been found to
develop a spin gap for $J^{\prime}/J \sim (t^{\prime}/t)^2 > 0.2412$ 
\cite{Haldane_82,Eggert_96} and incommensurate antiferromagnetic order
for $J^\prime/J > 0.5$.\cite{WhiteAffleck_96} 
This picture has been
confirmed numerically. \cite{Fabrizio_98,DaulNoack_00}

\begin{figure}[tbh]
\includegraphics[width=\smallfig]{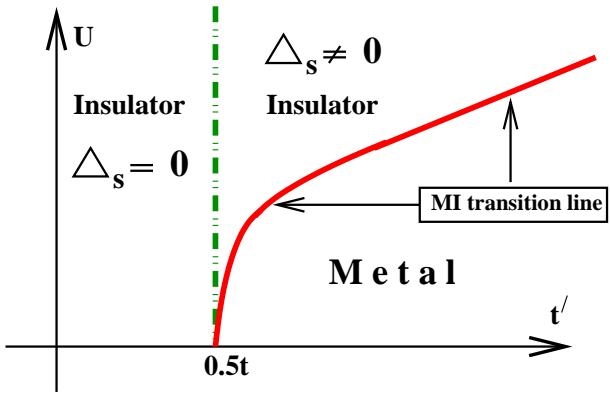}
\caption[DIAGONAL]{Qualitative sketch of the ground state phase diagram of
the half-filled $t-t^{\prime}$ model. The solid line marks the
metal-insulator transition. The dashed line indicates the 
transition from a gapless spin excitation spectrum at $t^{\prime} \leq 0.5t$
to the spin-gapped phase. }
\label{Fig:Fig3_PhaseDiagram.eps}
\end{figure}

For $t^{\prime}> 0.5t$, the Fermi level intersects the one-electron band at 
four points $\left(\pm k_{F}^{\pm}\right)$. 
This is the origin of more complex behavior for
weak and intermediate values of $U$.
For weak coupling ($U \ll t$), the ground-state phase diagram is well
understood in the  
two-chain limit ($t^{\prime}\gg t$).\cite{Fabrizio_96} In this
case, the Fermi vectors $k_{F}^{\pm}$ are sufficiently far from
$\pi/2$ to suppress 
first-order umklapp processes. Therefore, the system is metallic. 
The infrared behavior 
is governed by the low-energy excitations in the vicinity of the four
Fermi points,  
in full analogy with the two-leg Hubbard model.\cite{BalentsFisher_96}
Thus, while the charge excitations are gapless, the spin degrees of
freedom are gapped.
\cite{BalentsFisher_96,Fabrizio_96,DaulNoack_00,Gros_01,Torio_03}
Higher-order Umklapp processes become relevant for intermediate values of $U$
because  the Fermi momenta fulfill the condition $4(k_{F}^{+}-k_{F}^{-})=2\pi$
(at half-filling).  Therefore, starting from a metallic region for small $U$
at a given value of $t'$ ($t'> 0.5 t$), one reaches a transition line 
$U=U_{c}(t^{\prime})$,
above which the system is insulating with both charge and spin gaps.
\cite{Fabrizio_96}
The gross features of the phase
diagram are depicted in Fig. \ref{Fig:Fig3_PhaseDiagram.eps}.


\section{\bf Bosonization}
\subsection{Effective Theory}
We first consider the regime $U, t^{\prime} \ll t$ where
bosonization is applicable. 
We linearize the spectrum in the vicinity of the two Fermi points
$k_{F}=\pm \pi/2$ and go to the continuum limit by substituting
\be\label{C-expansion}
c_{n\sigma} \rightarrow  i^{n} \psi_{R\sigma}(x) +
(-i)^n \psi_{L\sigma}(x)\, ,
\ee
where 
the operators $\psi_{R\sigma} (x)$ and $\psi_{L\sigma} (x)$ are the
right and left components of the Fermi field, respectively.
These fields can be bosonized in a standard way,
\cite{GNT}
\begin{eqnarray}
\psi_{R\sigma} & \rightarrow & \frac{1}{\sqrt{2\pi\alpha}}
e^{ i\sqrt{4\pi}\phi_{R\sigma}}\nonumber\\
\psi_{L\sigma} & \rightarrow & \frac{1}{\sqrt{2\pi\alpha}}
e^{- i\sqrt{4\pi}\phi_{L\sigma}}\, \nonumber,
\end{eqnarray}
where $\phi_{R\sigma}$ ($\phi_{R\sigma}$) are the right(left)-moving
Bose fields and $\alpha$ is the infrared cutoff.
We define 
$\phi_\sigma=\phi_{R\sigma} + \phi_{L\sigma}$
and introduce linear combinations, 
$\varphi_{c}=(\phi_\uparrow + \phi_\downarrow)\sqrt{2}$ and 
$\varphi_{s}=(\phi_\uparrow - \phi_\downarrow) \sqrt{2}$, to describe
the charge and spin degrees of freedom, respectively.
Correspondingly, we introduce the conjugate fields
$\theta_\sigma=\phi_{L\sigma} - \phi_{R\sigma}$ and 
$\vartheta_{c} = (\theta_\uparrow + \theta_\downarrow)\sqrt{2}$ and 
$\vartheta_{s} = (\theta_\uparrow - \theta_\downarrow)\sqrt{2}$.
After a simple rescaling, we arrive at the bosonized
version of the Hamiltonian (\ref{t1t2Umodel})
$$
{\cal H} = {\cal H}_{s} + {\cal H}_{c}\, ,
$$
where both the spin part
\bea
{\cal H}_{s}&= v_{s} \int dx&\Big\{\frac{1}{2}(\partial_{x}\varphi_{s})^2
+\frac{1}{2}(\partial_x \vartheta_{s})^2\nonumber \\
&&+ \frac{m^{0}_{s}}{2\pi \alpha^2}\cos(\sqrt{8\pi}\varphi_{s}) \Big\},
\eea
and the charge part, 
\bea
{\cal H}_{c}& =  v_{c} \int dx&\Big\{\frac{1}{2K_{c}}(\partial_{x}\varphi_{c})^2
+\frac{K_{c}}{2} (\partial_x \vartheta_{c})^2 \nonumber \\
&&+ \frac{m^{0}_{c}}{2\pi \alpha^2}\cos(\sqrt{8\pi}\varphi_{c}) \Big\},
\label{SGc}
\eea
are described by the massive sine-Gordon model, with parameters
\bea
v_s &\approx& v_c \approx v_F, \nonumber \\
\left(K_{c}-1\right)&= &-2m^{0}_s = 2 m^{0}_{c} \approx - U/\pi t\, .
\eea
There is an important difference between $H_s$ and $H_c$ due to
the different stiffness constants. 
In the spin sector with $K_s=1$, the system is in the weak-coupling limit
and scales to a Gaussian model with gapless spin
excitations. 
In the charge sector with $K_c<1$, the system is in the strong coupling
regime and the low-energy behavior is dominated by the cosine term. 
In the ground state, the field $\varphi_c$ is pinned at one of the
minima of the cosine term and, correspondingly, there is a finite
energy gap for charge excitations.

Let us now discuss what happens when $t'$ increases and reaches values
of the order of $t/2$, where two additional Fermi points appear in the
band structure. 
For spin degrees of freedom, new scattering channels appear
at $t'=t/2$, and the system scales to strong coupling. 
Therefore, a spin gap is expected to open for $t'>t/2$,
very much like in the case of two coupled Hubbard chains.
\cite{BalentsFisher_96}

For the charge degrees of freedom, the situation is more complicated (and
more interesting) because the charge gap blocks new scattering
channels until $t'$ is made sufficiently large so that additional
states emerge beyond the gapped region.
Thus, for $t'$ slightly above $t/2$, the Fermi momentum
changes without affecting the Umklapp processes. 
The Hamiltonian is still given by
Eq.~(\ref{SGc}), but in order to allow for a change of particle number
around the Fermi 
points, we have to add a topological (chemical potential) term 
\be
\delta H_c=-
\frac{\mu_{\text{eff}}}{\sqrt{2\pi}} 
\int dx\; \partial_{x}\varphi_{c}\, ,
\ee
where
\be
\mu_{\text{eff}} =  \left\{ 
\begin{array}{l}
    0 \hspace*{2.4cm} \mbox{for} \quad 
    t^{\prime} < 0.5t \\
    \frac{t^2}{2t^{\prime}}-2t^{\prime} \neq 0 \qquad 
    \mbox{for} \quad 
    t^{\prime} > 0.5t\, . \\ 
\end{array}
\right . 
\label{ChemPot_gg}
\ee
$H_c+\delta H_c$ is the standard Hamiltonian for the
commensurate-incommensurate transition,\cite{GNT,Giamarchi}
which has been intensively studied in the past
using bosonization \cite{C_IC_transition} and
the Bethe ansatz.\cite{JNW_1984}

\subsection{Commensurate-incommensurate transition}

We now apply the theory of commensurate-incommensurate transitions to
the metal-insulator transition as a function of $t^{\prime}$.
At $\mu_{\reff}=0$ and $K_{c}<1$, the ground state of the 
field $\varphi_{c}$ is pinned at
\be
\langle 0| \sqrt{8\pi}\varphi_{c}|0 \rangle_{0} = 2\pi n \, .
\label{minima}
\ee
 The presence of the effective chemical potential makes it necessary to consider the ground state of the sine-Gordon model in sectors with nonzero topological charge. 
Using the standard expression for charge density 
in the case of two Fermi points,\cite{GNT}
\bea
\rho_{c}(x) &\simeq& \frac{1}{\sqrt{2\pi}}\partial_{x}\varphi_{c} \nonumber\\
&+&
A_{2k_{F}}\cos(2k_{F}x)\sin(\sqrt{2\pi}\varphi_{c})
\cos(\sqrt{2\pi}\varphi_{s})\nonumber\\  
&+& A_{4k_{F}}\cos(4k_{F}x)\cos(\sqrt{8\pi}\varphi_{c}) \, ,
\label{Charge_density_HubModel}
\eea
we observe that the pinning of the field $\varphi_{c}$ in one of the minima (\ref{minima}) suppresses the $2k_{F}$ charge fluctuations and stabilizes the $4k_{F}$ component. Any distortion of the $4 k_F$ charge distribution would require
an energy greater than the charge gap. This competition between the chemical potential term and commensurability drives a continuous phase transition from a gapped (insulating) phase at $\mu_{\reff} < \mu_{\reff}^{c}$ to a gapless (metallic) phase at
\be\label{CriticalLine}
\mu_{\reff} > \mu_{\reff}^{c} = \Delta_{c}\, ,
\ee
where $\Delta_{c}$ is the charge gap at $\mu_{\reff} = 0$.
\begin{figure}[tbh]
\includegraphics[width=\verysmallfig]{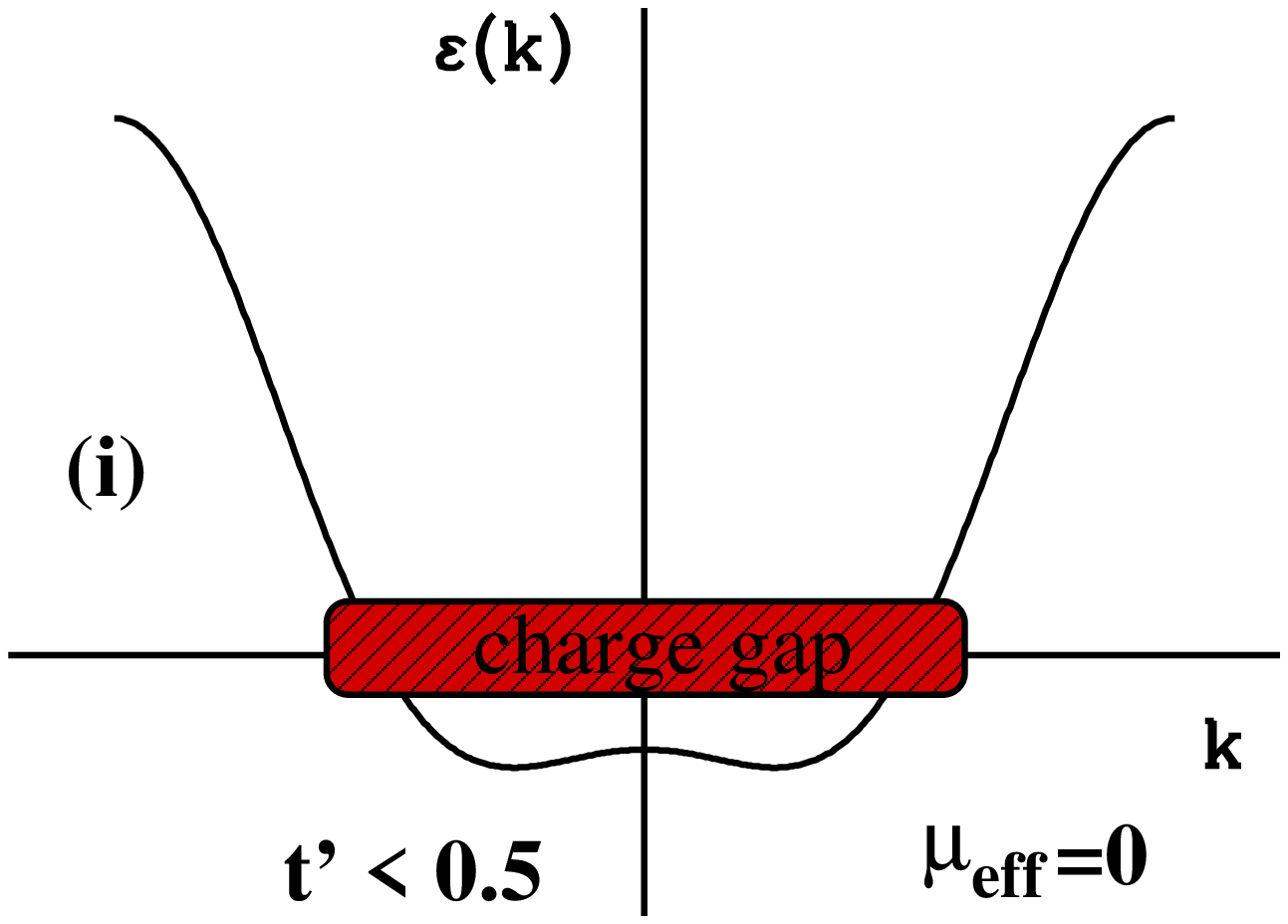}\\[0.3cm]
\includegraphics[width=\verysmallfig]{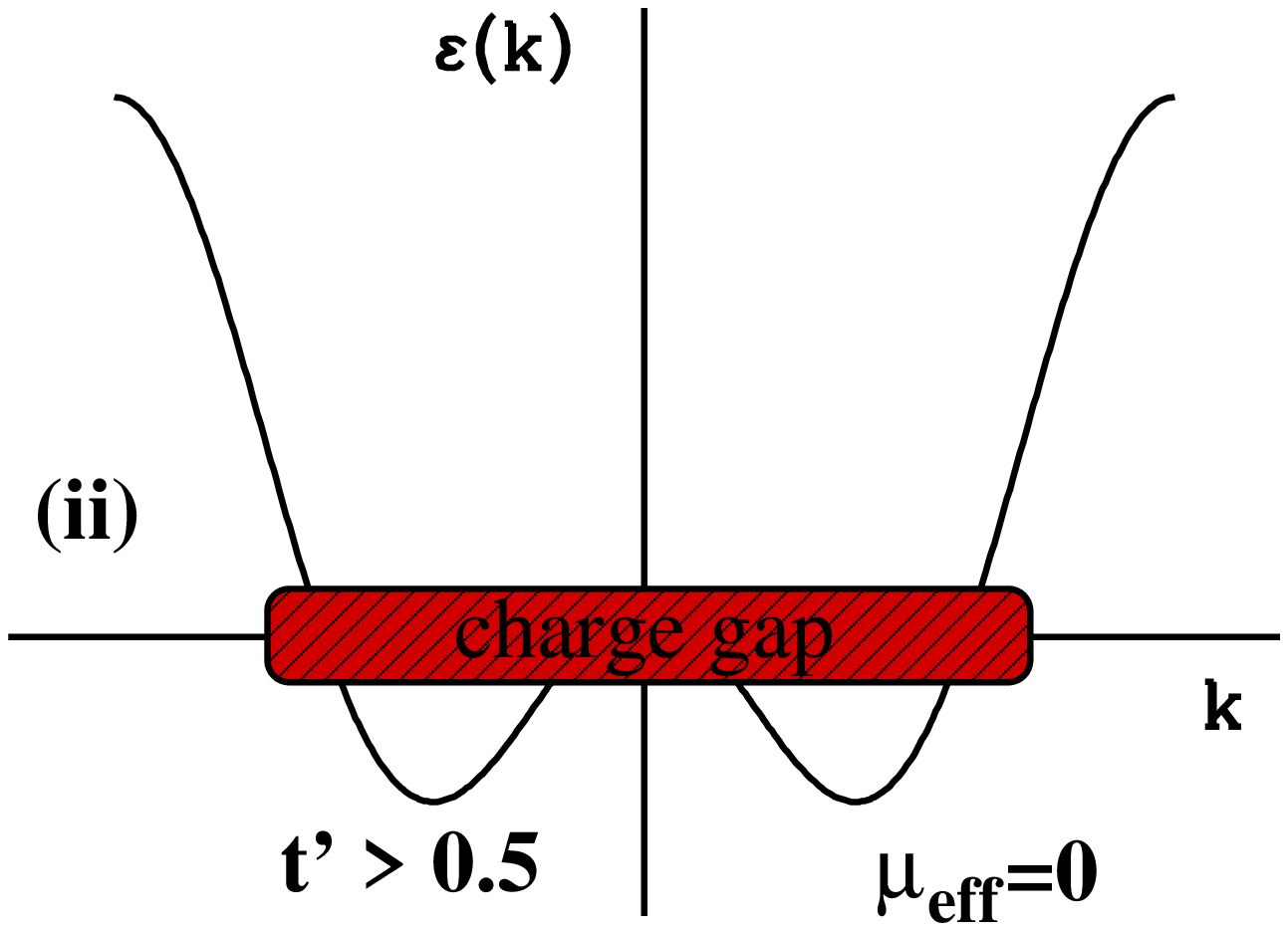}\\[0.3cm]
\includegraphics[width=\verysmallfig]{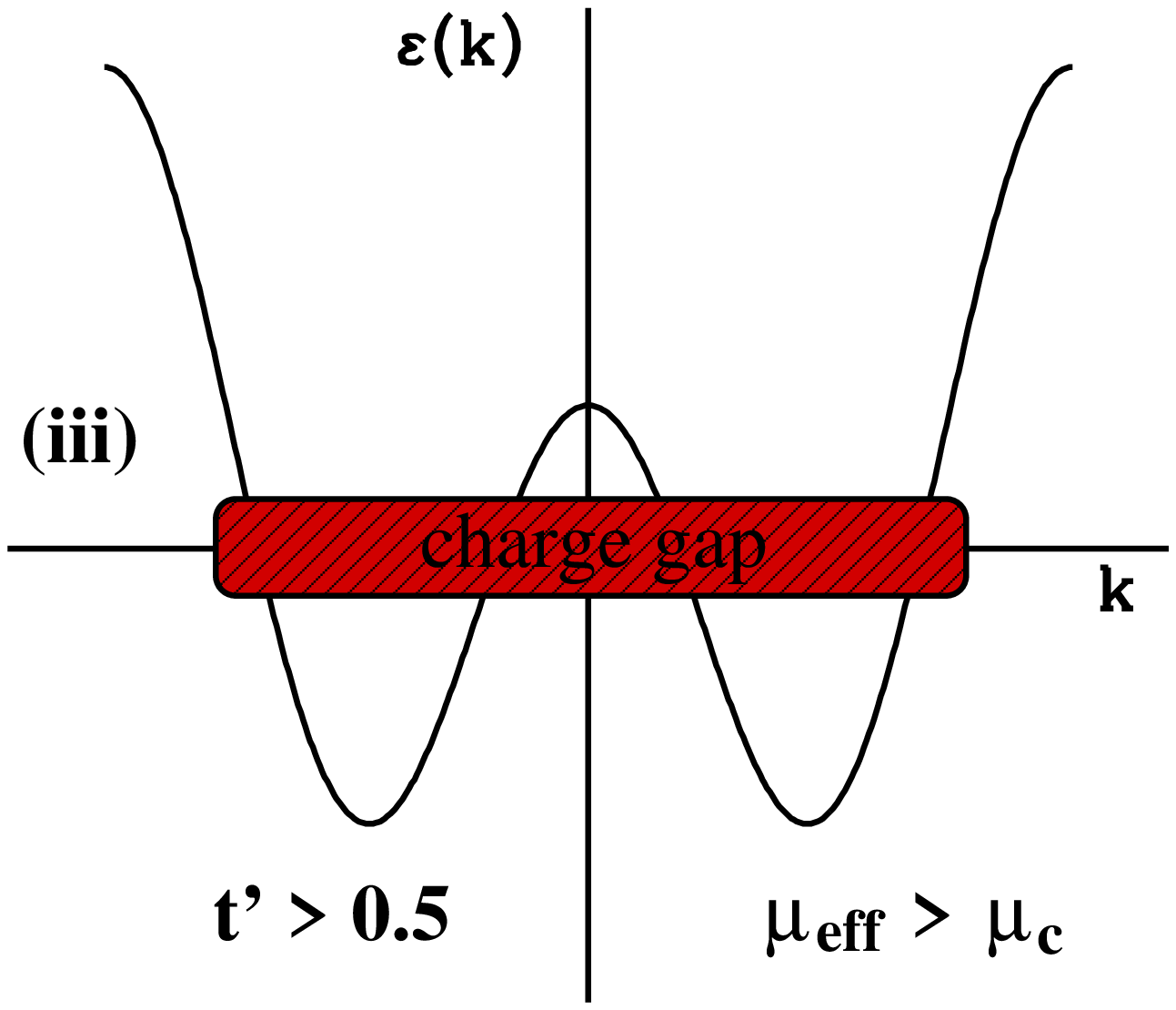}
\caption[DIAGONAL]{Sketch of the energy dispersions for the three
  regimes {\it (i)}, {\it (ii)}, and {\it (iii)}.
}
\label{Fig:Fig4-EnegyDispersion.eps}
\vspace{5mm}
\end{figure}

We now separately consider the qualitative behavior of the system in
the following three parameter regimes: {\it (i)} $t^{\prime} < 0.5t$,
{\it (ii)} 
$0.5t < t^{\prime} < t^{\prime}_{c}$, and {\it (iii)}
$t^{\prime} > t^{\prime}_c$, illustrated in
Fig.\ref{Fig:Fig4-EnegyDispersion.eps}.
In regime {\it (i)}, $t^{\prime} < 0.5t$, we expect a charge gap
$\Delta_c(U,t^{\prime}) \approx \Delta_c(U,t^{\prime}=0)$ and no spin
gap, as in the simple Hubbard model ($t^{\prime} = 0$).
In regime {\it (ii)}, $0.5t < t^{\prime} < t^{\prime}_c$, the spin gap opens
while the charge gap is reduced as \cite{note_1}
\be
\Delta_{c}(U,t^{\prime}) = \Delta_{c}(U,0.5t)-\mu_{\reff} \, ,
\label{eq:gapmueff}
\ee
where $\mu_{\reff}$ is given by Eq.\ (\ref{ChemPot_gg}).
Therefore, the charge gap decreases with increasing $t^{\prime}$ and
tends to zero at a $t_c^\prime$ qualitatively given by
\be
\Delta_{c}(U,0) - 2t^{\prime}_c + t^{2}/2t^{\prime}_c = 0 \, .
\label{CriticalLine2}
\ee
In regime {\it (iii)}, $t^{\prime} > t^{\prime}_c$,
the behavior of the system is characterized by
four Fermi points, $\pm k_F^{\pm}$.
The charge excitations are gapless, while the spin excitations are,
generically, gapped. \cite{Fabrizio_96,BalentsFisher_96}
Charge fluctuations will be characterized by two dominant periodic
modulations with wave vectors
$2 k_F^-$ and $2 k_F^+$.
For $t^\prime$ slightly larger than $t^\prime_c$, the usual
charge-density wave ($2 k_F^+\approx \pi$) is accompanied by
a long-wavelength modulation at 
\be\label{Q*}
2 k_F^- = \sqrt{(t^{\prime}-t^{\prime}_{c})/t^{\prime}_{c}}.
\ee

\begin{figure}[tbh]
\begin{center}
\includegraphics[width=\verysmallfig]{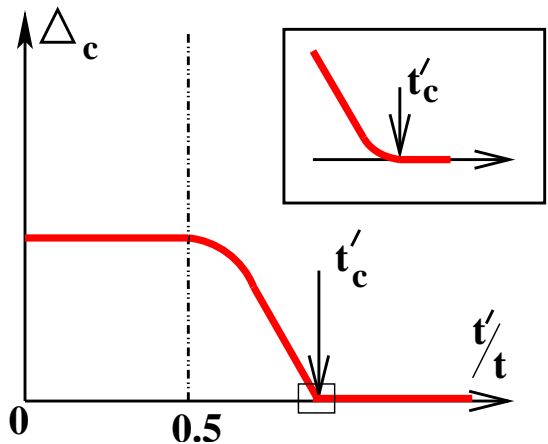}
\caption[DIAGONAL]{Sketch of the
charge gap as a function of the parameter $t^{\prime}$. The inset
shows an enlargement of the vicinity of the transition point.
}
\label{Fig:Fig5_t1t2U_paper.eps}
\end{center}
\end{figure}

As soon as the new set of states in the vicinity of the Fermi
points $\pm k_F^-$ appears, a channel for higher order umklapp
scattering processes opens.
In the sector of the phase
diagram characterized by two-band behavior ($t^\prime \gg t$),
these processes are responsible for the opening of a charge gap at the
transition line.\cite{Fabrizio_96}
Therefore, in the parameter range where the renormalized one-band
(Hubbard) gap (\ref{eq:gapmueff}) becomes exponentially small, a
crossover to the regime of two-band behavior takes
place.
Therefore, the linear decay of the charge gap as a function of 
$t^\prime$ crosses over to exponential behavior. 
The evolution of the charge gap as a function of $t^\prime$
is sketched in Fig.\ \ref{Fig:Fig5_t1t2U_paper.eps}.

\subsection{Two-chain limit}

To conclude our analysis, we discuss the limit of strong
next-nearest-neighbor hopping ($t^{\prime}\gg t$). 
For $t=0$, the system
is decoupled into two half-filled Hubbard chains and, for arbitrary
$U>0$, the ground state corresponds to a Mott insulator.  
The origin of the insulating behavior is the commensurability of umklapp
scattering between the Fermi points, located at $\pm \pi/4$ and 
$\pm 3\pi/4$.  
When $t \neq 0$ this commensurability is lost.  
The Fermi points are shifted with respect to their values at $t=0$, and the
Fermi energy (the chemical potential for $U=0$) moves away from $0$
to $\varepsilon_F\approx -t^2/{2t'}$ (for $t\ll t'$). 
For large enough values of $t$, the system is therefore expected to be
metallic.

\begin{figure}[tbh]
\includegraphics[width=\verysmallfig]{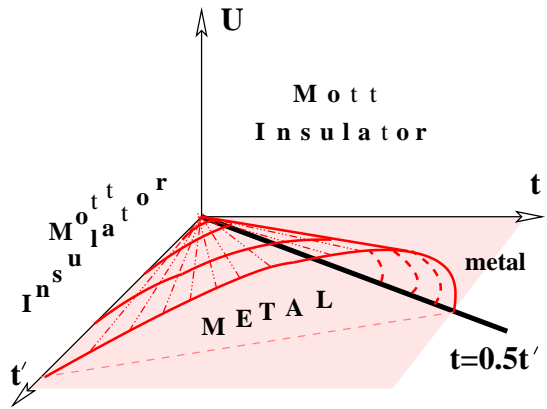}
\caption[DIAGONAL]{Qualitative 
phase diagram of the half-filled repulsive $t-t^{\prime}$ Hubbard
chain. A gapless charge excitation spectrum (metallic phase) 
exists at $U=0$ for arbitrary $t$ and $t^{\prime}$ and for $U>0$
in the sector of parameter space below the ``roof'' covering the
area $U<U_{c}$ between the lines $t^{\prime}=0.5t$ and $t=0$ in
the $U=0$ plane.} \label{Fig:t1t2U_paper_Fig_PD_FULL.eps}
\end{figure}

In order to estimate the location of the Mott transition, we can use a
similar argument to the one given above for $t' \approx 0.5$. 
As long as the chemical potential is smaller than
the charge gap, the system remains an insulator.
A transition to a metallic phase is expected to occur for
$\varepsilon_F$ of the order of $\Delta_c$, i.e., 
for $t^2\approx (Ut'^3)^{\frac{1}{2}}\exp (-2\pi t'/U)$.  
A qualitative sketch of the phase diagram is given in Fig.
\ref{Fig:t1t2U_paper_Fig_PD_FULL.eps}.

\section{Numerical results}

In order to investigate the detailed behavior of the metal-insulator
transition and to test the validity of the picture obtained from
bosonization, we have carried out numerical calculations using the
DMRG.\cite{White_92}
We have calculated the properties of the ground state and low-lying
excited states for systems with open boundary conditions of lengths
between $L=32$ and $L=128$ sites, keeping up to $m=1000$
density-matrix eigenstates.
As we shall see in the following, the finite-size effects are
quite large in certain parameter regimes, so that a careful
finite-size scaling must be carried out.


\subsection{Transition line}

\begin{figure}[tbh]
\includegraphics[width=\smallfig]{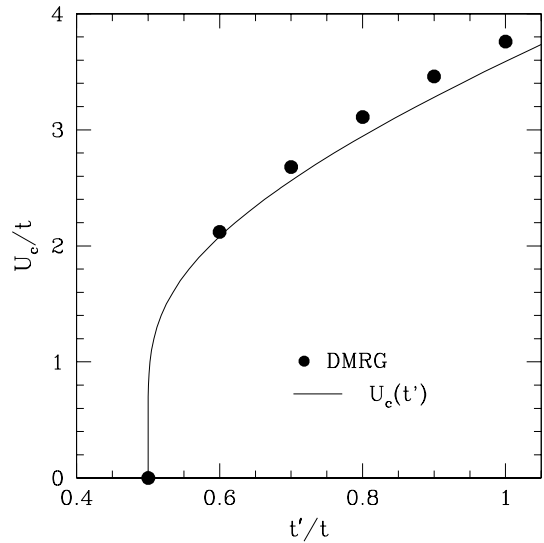}
\caption[DIAGONAL]{The metal-insulator transition line in the $t-t^{\prime}-U$
model with $t=1$ obtained from DMRG studies
\cite{AebBaerNoack_01,aebischer_thesis} 
(black circles) and from Eq.\ (\ref{CriticalLine2}) (solid line).
}
\label{Fig:PhaseBoundary_ChargeGap}
\end{figure}

The critical behavior of the metal-insulator transition as a function
of $U/t$ for $t^\prime > 0.5 t$ can be obtained from the behavior of the
electric susceptibility, which diverges in going from an insulator to
a metal. \cite{AebBaerNoack_01,aebischer_thesis}
In Fig.\ \ref{Fig:PhaseBoundary_ChargeGap}, we display the
transition line in the $t-t^{\prime}-U$ model at $t=1$
obtained from the DMRG \cite{AebBaerNoack_01,aebischer_thesis}
and from Eq.\ (\ref{CriticalLine2}). 
The agreement between the DMRG results and  Eq.\ (\ref{CriticalLine2}) is
remarkably good.


\subsection{\bf Spin and charge gaps}

In order to investigate the predictions of the continuum theory, we
calculate the charge gap, defined as
\be
 \Delta_{c} = \frac{1}{2} \left[ E_0(N+2,0) + E_0(N-2,0)
  -2E_0(N,0) \right]
\ee
and the spin gap, 
\be
 \Delta_{s} = E_0(N,1) - E_0(N,0)\,  ,
\ee
where $E_0(N,S)$ is the ground-state energy for $N$
particles and spin $S$ on a chain of fixed length $L$, using the DMRG.

\begin{figure}[tbh]
\includegraphics[width=\smallfig]{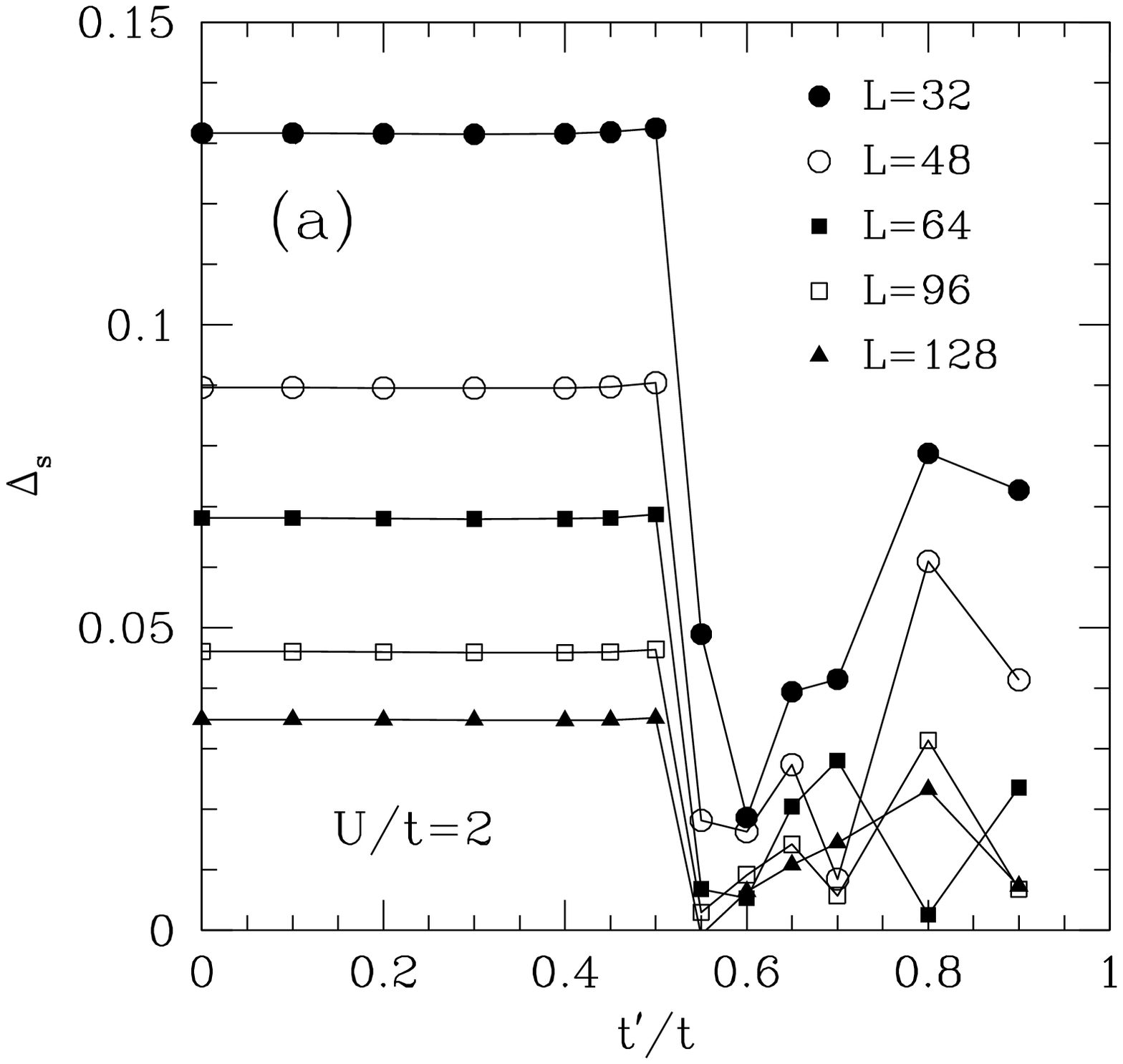}
\includegraphics[width=\smallfig]{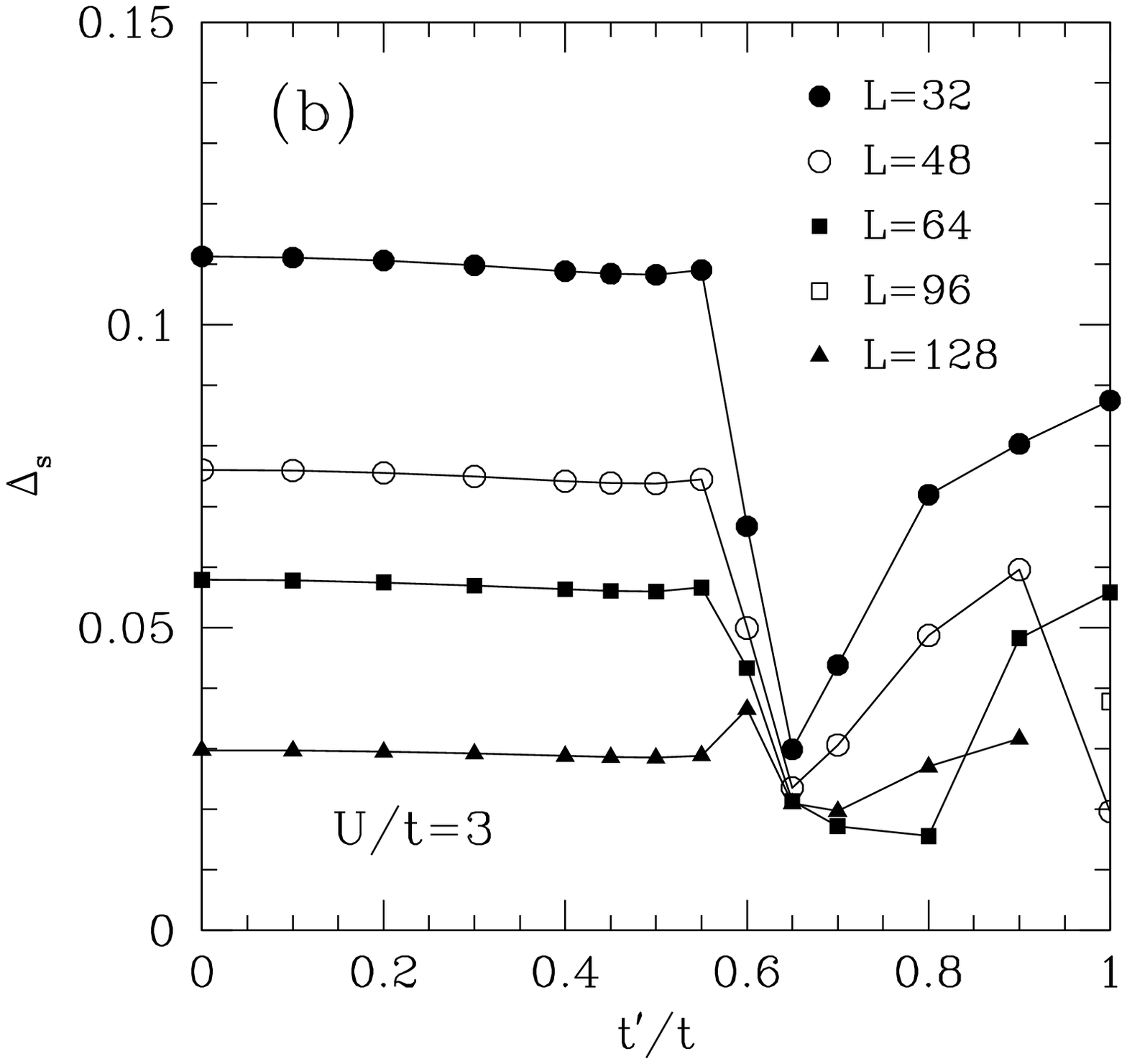}
\caption[DIAGONAL]{Spin gap of as a function of $t^{\prime}$
for (a) $U/t=2$ and (b) $U/t=3$.
}
\label{Fig:SpinGap}
\end{figure}

\begin{figure}[tbh]
\includegraphics[width=\smallfig]{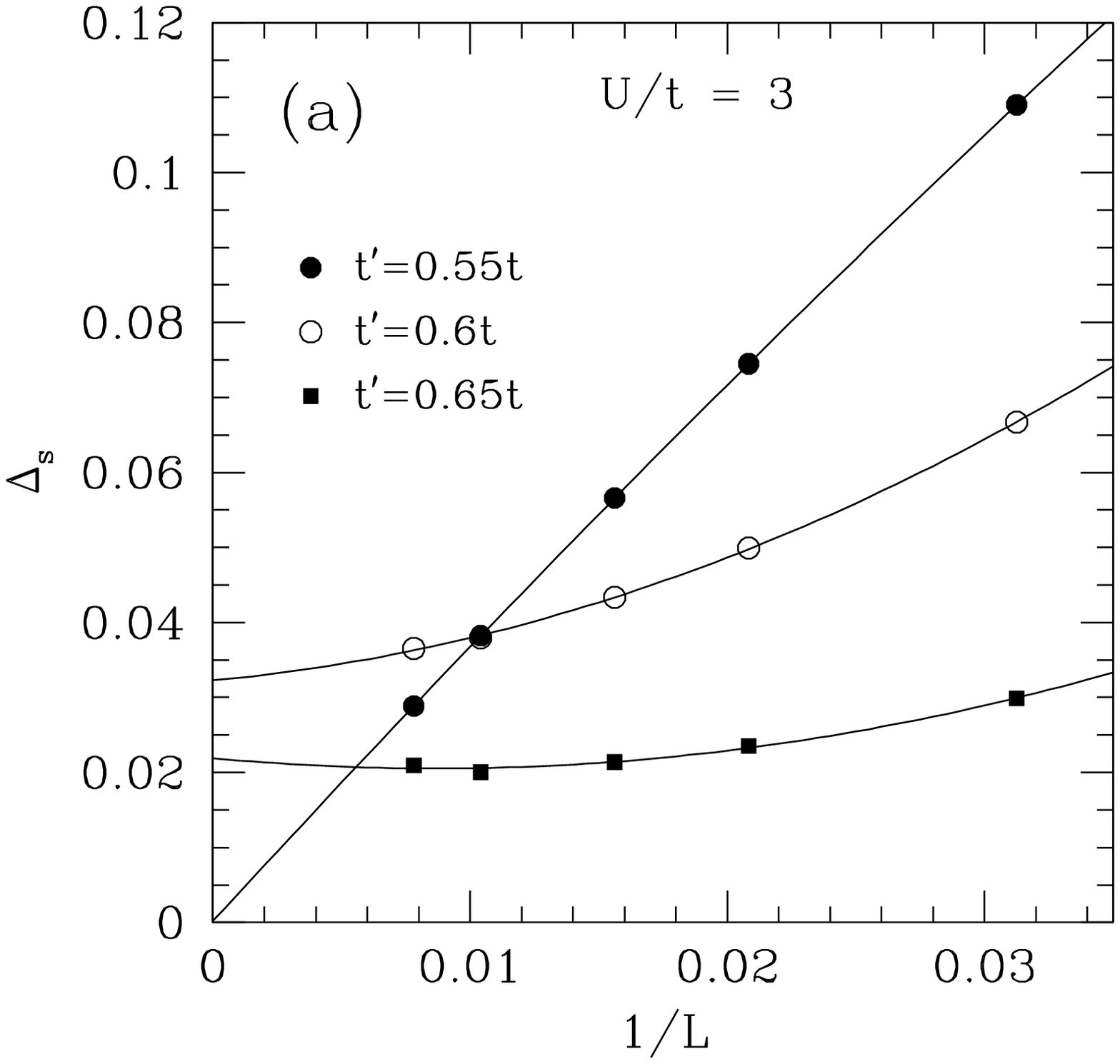}
\includegraphics[width=\smallfig]{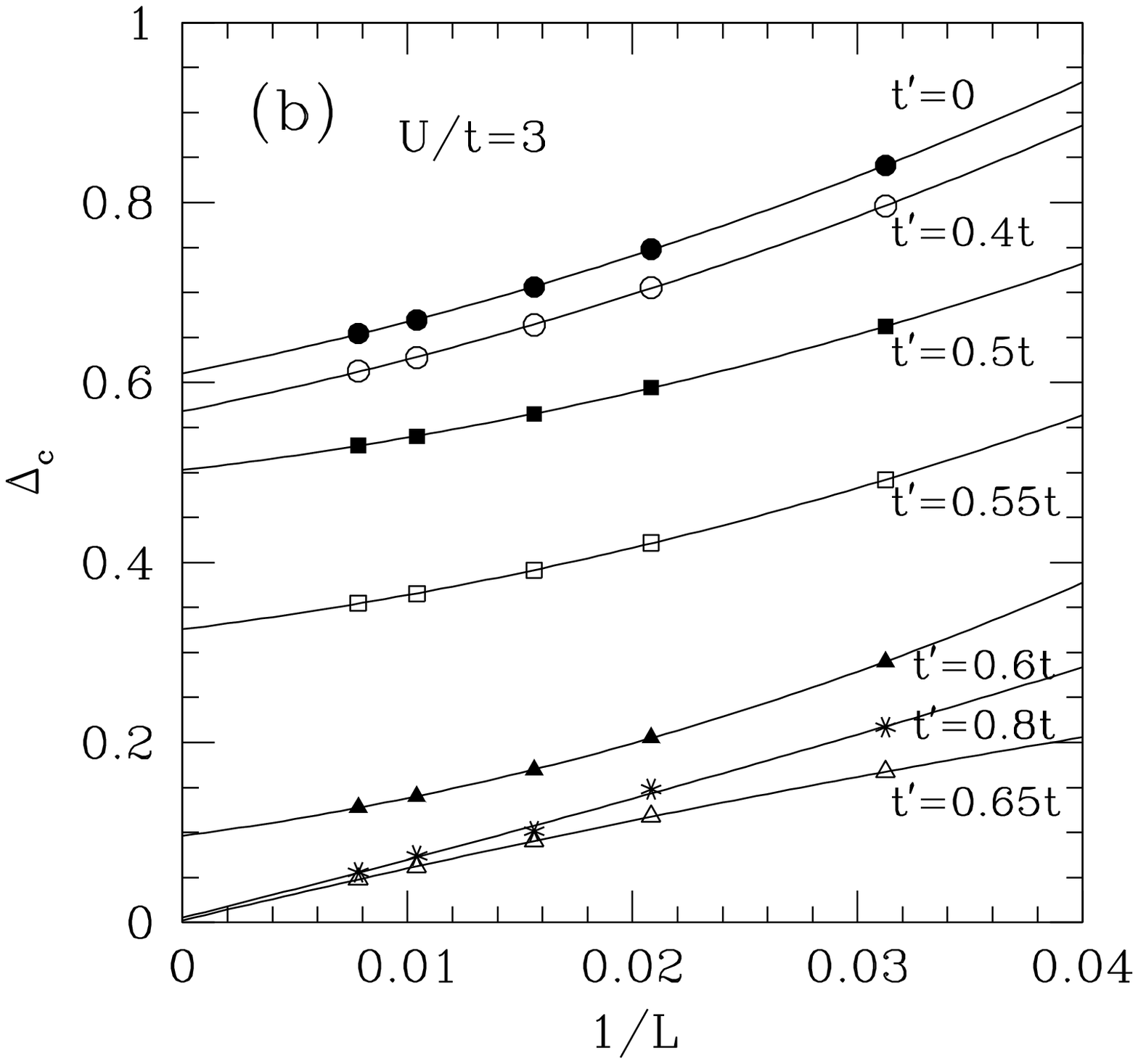}
\caption[DIAGONAL]{(a) Spin gap and (b) charge gap of as a function of
  $1/L$ for $U/t = 3$ and various values of $t'/t$.
}
\label{Fig:GapsLi}
\end{figure}

In Fig.\ \ref{Fig:SpinGap}, we display the spin gap as a function of
$t^{\prime}$ at $U/t=2$ and $U/t=3$ 
for various values of the chain length $L$. 
As can be clearly seen, for $0< t^{\prime} \leq 0.5t$ the spin
excitation spectrum on finite chains does not depend on 
$t^{\prime}$.
For $t^{\prime} \leq 0.5t$, the value of the
spin gap is found to coincide
with that of the half-filled Hubbard model ($t^{\prime} = 0$) which
vanishes in the infinite-chain limit [see Fig.\ \ref{Fig:GapsLi} (a)].  

A clear change in the $t^{\prime}$--dependence of the spin gap at $U/t=2$
takes place at $t^{\prime}=0.5t$, indicating the development of a new
phase in the spin sector. 
It is known from other studies 
\cite{BalentsFisher_96,Fabrizio_96,DaulNoack_00,Gros_01,Torio_03}
that a spin gap opens at a critical value of $t^{\prime}$ which is
is approximately at or slightly above $t^{\prime} =0.5t$, becoming
weakly larger at intermediate $U$ values.

In Fig.\ \ref{Fig:GapsLi}(a), we display the spin gap plotted as a
function of the inverse chain length for three values of $t^\prime$
near the transition at $U/t=3$.
At $t^\prime = 0.55 t$, the spin gap clearly scales to zero at
infinite system size, with the values at a particular system size
virtually identical to the $t^\prime = 0$ case and the scaling
predominantly linear in inverse system size.
For $t^\prime = 0.6 t$ and $t^\prime = 0.65 t$, the dominant scaling
term is quadratic rather than linear in $1/L$ and there is clearly
scaling to a finite value of the gap.
For $t^\prime = 0.65 t$, the size of the extrapolated gap is smaller 
than for  $t^\prime = 0.6 t$, and there is a slight upturn in the gap
at the largest system size, which, however, is not significantly
larger than the estimated error of the DMRG calculation, approximately
the symbol size.
However, for larger values of $t^\prime$, the finite-size behavior
becomes less regular, as can be seen in Figs.\ \ref{Fig:SpinGap}(a) and (b).
This behavior is due to the appearance of an incommensurate wave vector
characterizing the spin excitations that occurs when a substantial
density of states at all four Fermi points develops and makes it
virtually impossible to carry out a well-controlled finite-size
scaling for larger values of $t^\prime$.

The transition associated with the opening of the spin gap 
is independent from the insulator-metal
transition, as can be  clearly seen for $U/t=3$ (where the effect of
fluctuations is reduced). 
As is shown in Fig.\ \ref{Fig:SpinGap}(b) and 
Fig.\ \ref{Fig:GapsLi}(a), the spin gap opens for 
$t^{\prime}_{s} \geq 0.55t$, while the insulator-metal transition
takes place at $t^{\prime}_{c} \simeq 0.65t$ (see below). 
Note that the critical value of the
next-nearest-neighbor hopping amplitude, corresponding to an opening
of the spin 
gap at $U/t=3$, $t^{\prime}_{s} \geq 0.55t$ deviates from the line
$t^{\prime}_{s} \geq 0.5t$. 
Our findings agree with previous studies.
\cite{BalentsFisher_96,Fabrizio_96,DaulNoack_00,Gros_01,Torio_03}

A plot of the finite-size extrapolation of the charge
gap is displayed in Fig.\ \ref{Fig:GapsLi}(b) for various values of
$t^\prime$.
As can be seen, the behavior is well-behaved for values of $t^\prime$
from 0 to 0.8.
For $0\le t^\prime \le 0.6t$, the scaling has a substantial
positive quadratic term in $1/L$  and the gap is finite.
For $t^\prime=0.65 t$ and $0.8 t$, the extrapolated gap clearly vanishes and
there is a negligible or negative quadratic contribution.
For $t^\prime > 0.8$, the finite-size effects become irregular due to
incommensurability of the charge excitations, and finite-size
extrapolation becomes difficult.

\begin{figure}[tbh]
\includegraphics[width=\smallfig]{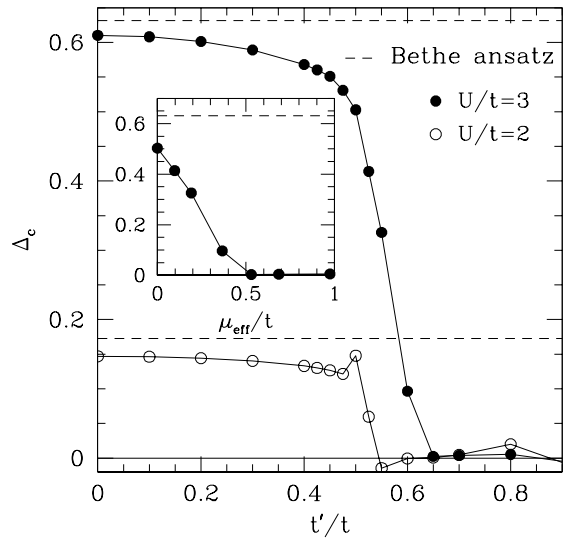}
\caption[DIAGONAL]{Charge gap of as a function of $t^{\prime}$ for
$U/t=3$ (black circles) and $U/t=2$ (open circles). 
The inset shows the charge gap as a
function of the parameter $\mu_{\reff}$ for $0.5<t^{\prime}<0.85t$.
}
\label{Fig:ChargeGap}
\end{figure}

In Fig.\ \ref{Fig:ChargeGap}, the $L=\infty$ extrapolated
value of the charge gap is displayed as a function of $t^{\prime}$ for
$U/t= 2$ and $U/t=3$.  
There is a clearly defined insulator-metal transition at $t_c = 0.55t$  at 
$U/t=2$ and $t_c = 0.65t $ for  $U/t=3$. 
Note that the charge gap goes smoothly to zero above $t^\prime = 0.5
t$ for $U/t=3$.
The inset in Fig.\ \ref{Fig:ChargeGap}
shows the charge gap for $U/t =3$  as a function of the parameter
$\mu_{\reff}=2t^{\prime}-t^{2}/2t^{\prime}$
for $0.5t<t^{\prime}<0.85t$.
As can be seen, the charge gap drops off approximately linearly with
$\mu_{\text{eff}}$, in agreement with Eq.\ \ref{eq:gapmueff}.
For $U/t = 2$, there is a somewhat irregular behavior of the charge
gap near the $t^\prime=0.5 t$.
In particular, there is a small peak exactly at $t^\prime=0.5$.
The finite-size scaling for this point is completely regular, however,
and we estimate the size of the total error, due to both the
extrapolation and the DMRG accuracy, to be less than the symbol size.
Therefore, in our estimation, the peak at $t^\prime=0.5$ is a real
effect.
For $t^\prime=0.55$, the value of the extrapolated charge gap is
slightly below zero.
This is due to errors in the finite-size extrapolation due to slightly
irregular behavior with system size.

\subsection{\bf Spin and charge densities}

\begin{figure}[tbh]
\includegraphics[width=\smallfig]{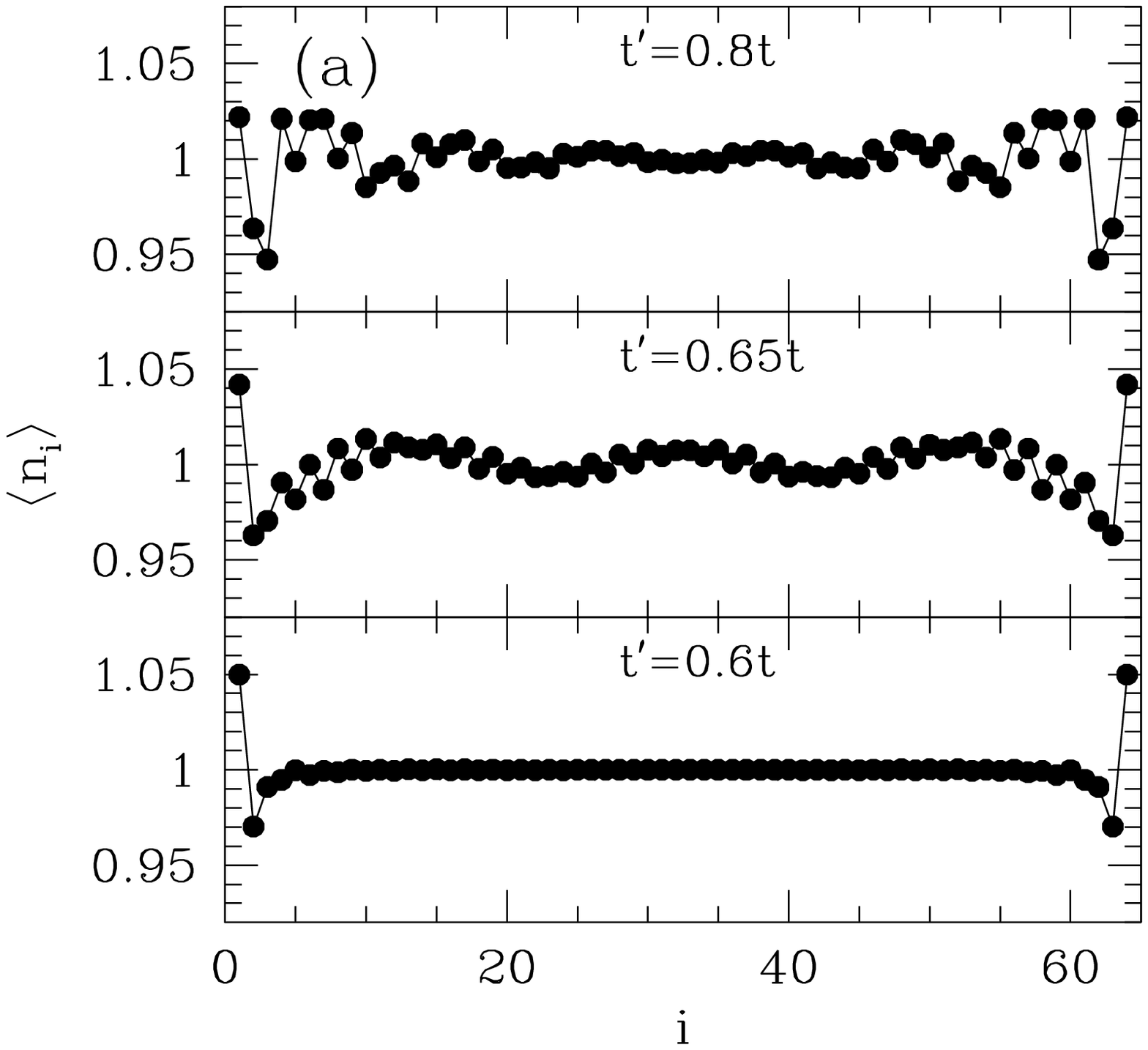}
\includegraphics[width=\smallfig]{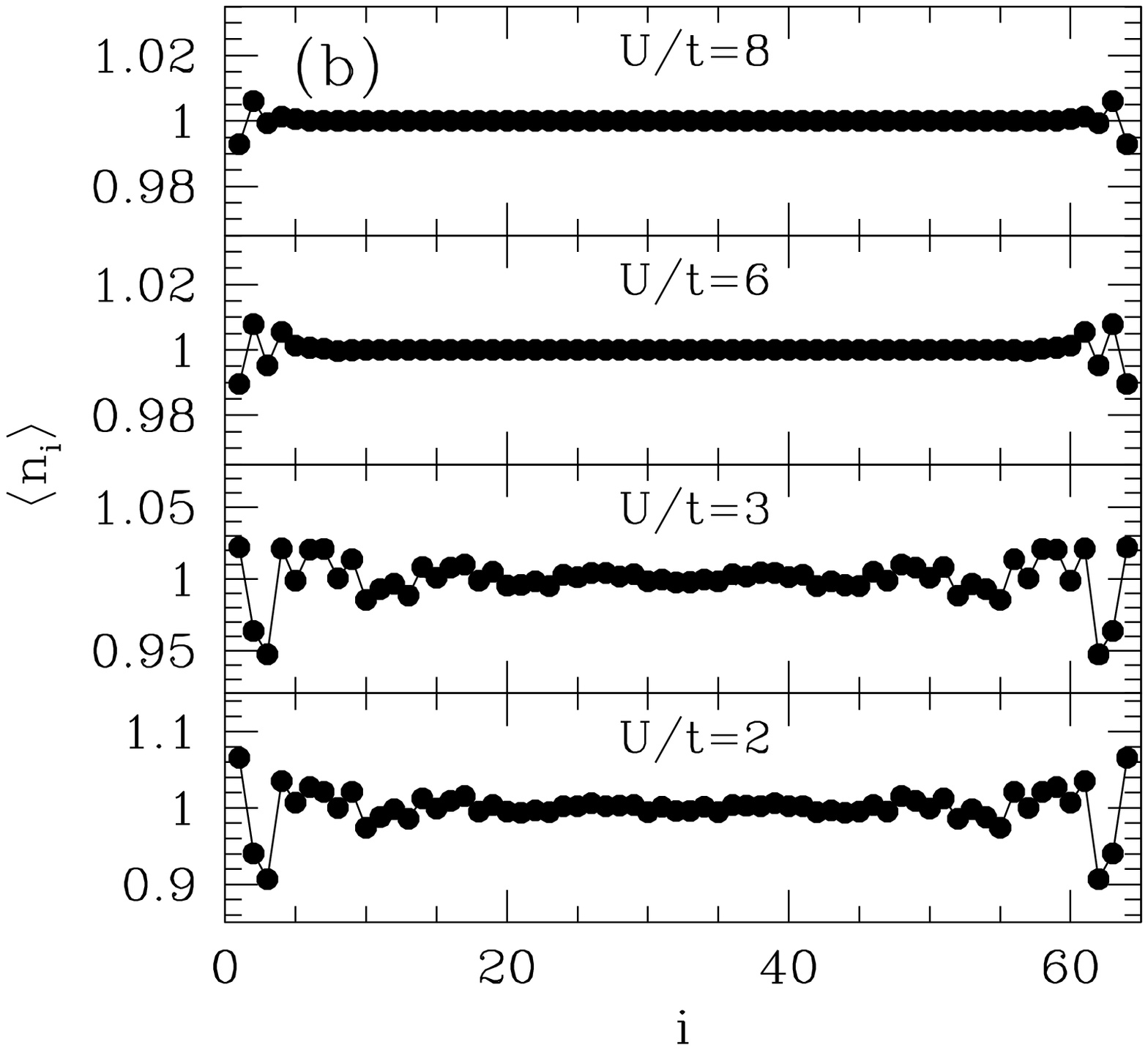}
\caption[DIAGONAL]{Charge distribution $\langle n_i\rangle$ in the
  ground state of the $t-t^{\prime}-U$ chain with (a) $L=64$, 
  $U/t=3$ and   
$t^{\prime}=0.6 t,\,0.65 t,\,0.8 t$ and (b) for  
$t^{\prime}=0.8t$  and  $U/t=2,\,3,\,6,\, 8$.}
\label{Fig:ChargeDensity_U_tp}
\end{figure}

Valuable insight into the nature of the insulator-metal transition can
be obtained by studying the charge and spin density distributions. 

We first examine the local charge density $\langle n_{i}\rangle$ in
the ground state  
with $L=64$ sites and $U/t=3$ as $t^{\prime}$
is varied. 
At $t^{\prime}=0.6t$, as  can be seen in 
Fig. \ref{Fig:ChargeDensity_U_tp}(a), 
the commensurate charge distribution characterizing the insulating phase 
(i.e., $\langle n_i\rangle = 1$) is reached within a few lattice sites
from the edge.  
The boundary effect is relatively weak and short-range.
The insulator-metal transition manifests itself at 
$t^{\prime}_{c} = 0.65t$ via the appearance of incommensurate
modulations in the charge distribution.  
One can clearly see the long-wavelength modulations 
in density originating from the opening of the Fermi surface for 
$\left| k \right|\leq K_{F}^{-}$. 
At $t^{\prime}=0.8t$,
the incommensurability increases and the presence of long wavelength
modulations of the charge density become more evident.

In Fig.\ \ref{Fig:ChargeDensity_U_tp}(b), we plot the charge
distribution 
at $t^{\prime}=0.8$ for various values $U$.
In the  metallic phase, at $U/t=2$ and $U/t=3$, the charge
distribution pattern is strongly incommensurate. 
At $U/t=6$ and $U/t=8$, the well-established commensurate pattern of the
charge distribution characterizes the insulating phase.

\begin{figure}[tbh]
\includegraphics[width=\smallfig]{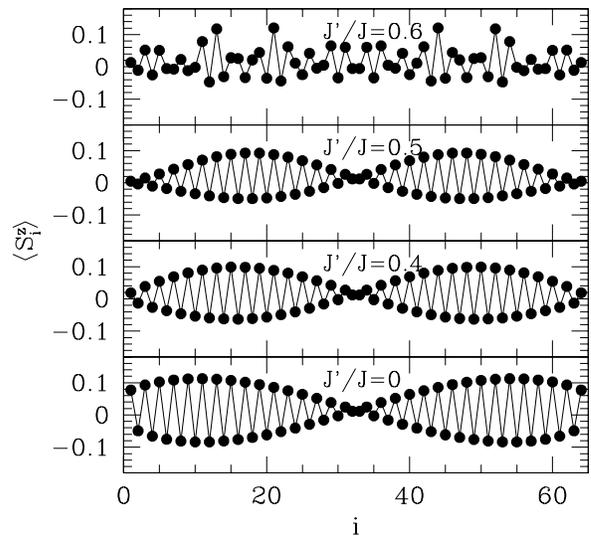}
\caption[DIAGONAL]{Spin distribution $\langle S^{z}_\rangle$ in the  $S^{z}_{\text{total}}=1$ state of the 
frustrated Heisenberg chain with $L=64$ and 
$J^{\prime}/J = 0, 0.4, 0.5$, and 0.6.}
\label{Fig:Multi_spinden_J1J2.eps}
\end{figure}

Let us now consider the spin density distribution. 
As has already been stressed above,  due to spin-charge
separation, the magnetic degrees of freedom in
the $t-t^{\prime}-U$ Hubbard model are not influenced by the
destruction of the ``holon Fermi surface'' caused by the dynamical
generation of a charge gap.
Instead, the development of the 
incommensurate spin distribution in the insulating phase is completely 
determined by the competition between the nearest-neighbor and
next-nearest-neighbor spin exchange interactions and reflects the itinerant
nature of the model only when the system is very close to the
insulator-metal transition or is in the metallic phase. 
To see this, we examine the
behavior of the spin distribution $\langle S^{z}_i\rangle$
in the $S^{z}_{tot}=1$ excited state of the $t-t^{\prime}-U$ Hubbard
model, comparing it to the effective model for strong coupling, the frustrated
Heisenberg chain.

We first examine the limiting case of the frustrated Heisenberg chain,
Hamiltonian (\ref{FH-chain}). 
In Fig.\ \ref{Fig:Multi_spinden_J1J2.eps}, we plot $\langle S^{z}_i\rangle$
calculated using the DMRG in the $S^{z}_{tot}=1$ state for various
values of the next-nearest-neighbor exchange $J^{\prime}$. 
At $J^{\prime}=0$, the unfrustrated case, there is one node in the
distribution of the spin 
density, which corresponds to two spatially separated spin-$S=1/2$
spinons.\cite{SAAP_1998}
These two spinons characterize the spin excitations up to the 
Majumdar-Ghosh point $J^{\prime}=0.5J$ ($t^{\prime} \simeq 0.7t$).
\cite{Majumdar_Ghosh_1969}
For larger values of the next-nearest-neighbor exchange,
the spin distribution becomes incommensurate.
\cite{WhiteAffleck_96}
In the plot for $J^\prime/J=0.6$, the absence of the commensurate
two-spinon structure of the excitations and the appearance of the
incommensurate wave vector can be seen. 

\begin{figure}[tbh]
\includegraphics[width=\smallfig]{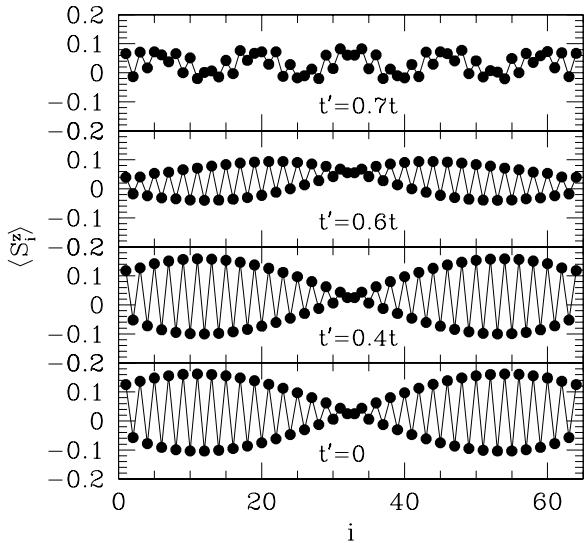}
\caption[DIAGONAL]{Spin distribution $\langle S^{z}_i\rangle$ 
in the  $S^{z}_{\text{total}}=1$ state of the 
$t-t^{\prime}-U$ chain with $L=64$, $U/t=3$ and
$t^{\prime} = 0, 0.4 t ,0.6 t$, and $0.7 t$.}
\label{Fig:Spin_Density_U3_tp.eps}
\end{figure}

\begin{figure}[tbh]
\includegraphics[width=\smallfig]{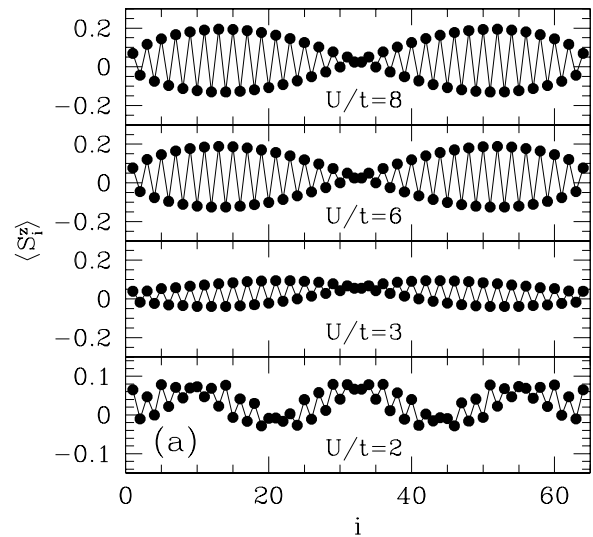}
\includegraphics[width=\smallfig]{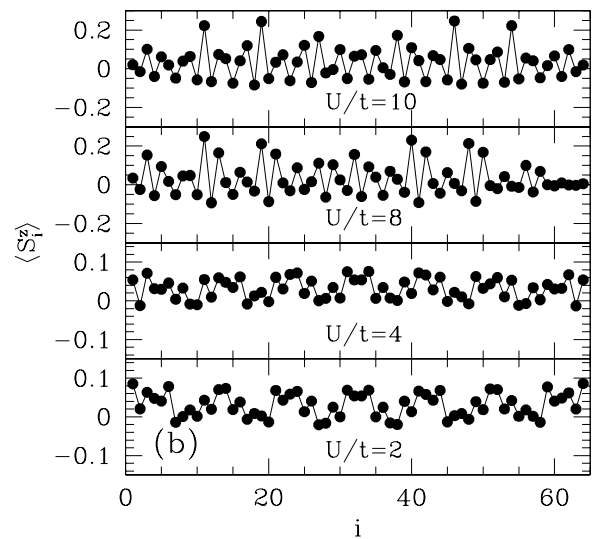}
\caption[DIAGONAL]{ Spin distribution $\langle S^{z}_i\rangle$ in the 
$S^{z}_{\text{total}}=1$ state of the 
$t-t^{\prime}-U$ chain with $L=64$ and for (a)
  $t^{\prime} = 0.6t$ with $U/t=2,3,6,8$ and (b) $t^{\prime} = 0.8 t$ with
  $U/t=2,4,8,10$.
}
\label{Fig:Multi_spinden_tp.eps}
\end{figure}

We next examine the behavior for the $t-t^{\prime}-U$ chain.
In Fig \ref{Fig:Spin_Density_U3_tp.eps}, we show the spin
distribution $\langle S^{z}_i\rangle$ in the  $S^{z}_{\text{total}}=1$ state with $L=64$,
$U/t=3$ and various values of $t^\prime$.
The single-node pattern characteristic of two spinons, as found in the
unfrustrated and weakly frustrated Heisenberg chain, can be seen for
$t^\prime = 0.6t$ and smaller.
At $t^{\prime} = 0.7 t$ an incommensurate pattern appears.
This is just above the value of $t^{\prime}$ at which the
charge gap goes to zero, $t^\prime_c \simeq 0.65 t$,
(see Fig.\ \ref{Fig:ChargeGap}) and the system becomes metallic.
However, for $t^{\prime}>0.7t$ the point at which the spin-density
distribution becomes incommensurate is independent of the insulator-to-metal 
transition.
To demonstrate this, we plot the spin
density distribution in the $S^{z}_{\text{total}}=1$ state for
increasing on-site repulsion $U$ for $t^{\prime} = 0.6t$ and 
$t^{\prime} = 0.8t$ in Fig. \ref{Fig:Multi_spinden_tp.eps}. 
As can be seen in  Fig. \ref{Fig:Multi_spinden_tp.eps}(a), for
$t^{\prime} = 0.6t$ the spin-density distribution is incommensurate in
the metallic phase at $U/t=2$, but 
acquires the  commensurate single-node pattern characteristic of two
spinons for on-site repulsions of $U/t=3$ and higher.
In contrast, for $t^{\prime} = 0.8t$,
Fig. \ref{Fig:Multi_spinden_tp.eps}(b), 
the incommensurate pattern of the spin distribution in the metallic phase
($U/t=2$) remains not only in the vicinity of the insulator-metal transition
at $U/t=3$ but also deep into the insulating phase at $U/t=8$ and $U/t=10$.

\subsection{Excitations for Large $t^{\prime}$}

We examine the behavior of the $t-t^{\prime}-U$ chain for large
next-nearest hopping ($t^{\prime}\gg t$), a limit which corresponds to
two chains coupled with a weak zigzag hopping.
In particular, we numerically investigate the transition from a two-chain 
(four-Fermi-point) metallic regime at weak $U$ to the
strong-coupling regime, for which the effective model is two
spin-$S=1/2$ Heisenberg chains coupled with a frustrating zigzag
interaction at $U \gg t^{\prime}\gg t$, i.e., $J^{\prime} \gg J$.

\begin{figure}[tbh]
\includegraphics[width=\smallfig]{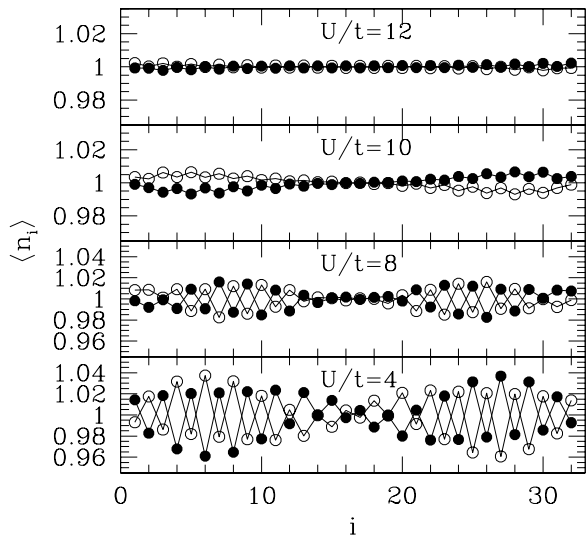}
\caption[DIAGONAL]{Charge distribution $\langle n_i\rangle$ in the
  ground state of 
the $t-t^{\prime}-U$ chain with $L=64$, $t^{\prime}=3 t$ and 
$U/t=4,8,10,20$. 
Black circles correspond to odd and open circles to
even lattice sites.
}
\label{Fig:2_Chain_CD}
\end{figure}

In Fig. \ref{Fig:2_Chain_CD}, we show the charge distribution 
$\langle n_i\rangle$
in the ground state of the $t-t^{\prime}-U$ chain for $t^{\prime}=3 t$,
$L=64$, and  $U/t=4, 8, 10$, and $20$. 
In the metallic phase ($U/t=4$ and $U/t=8$), the relative charge
density on the two chains alternates with relatively strong
amplitude and is modulated by an incommensurate wave vector.
As $U/t$ is increased to 10, the amplitude of the charge
fluctuations between the chains is strongly suppressed and the
incommensurate structure of the charge distribution changes to a very
distinctive two-bubble pattern.
This two-bubble distribution corresponds to two spatially separated
electron-hole pairs with an electron on even chain and a hole on odd
chain in the left part of the lattice and 
the inverse configuration in the right half of the lattice.
(Note that the ends break the symmetry between the two chains.)
Deeper into the insulating phase, at $U/t=20$, the charge density is
almost smooth and equal between the chains, except for a small
residual alternation of charge density due to the end effects.

\begin{figure}[tbh]
\includegraphics[width=\smallfig]{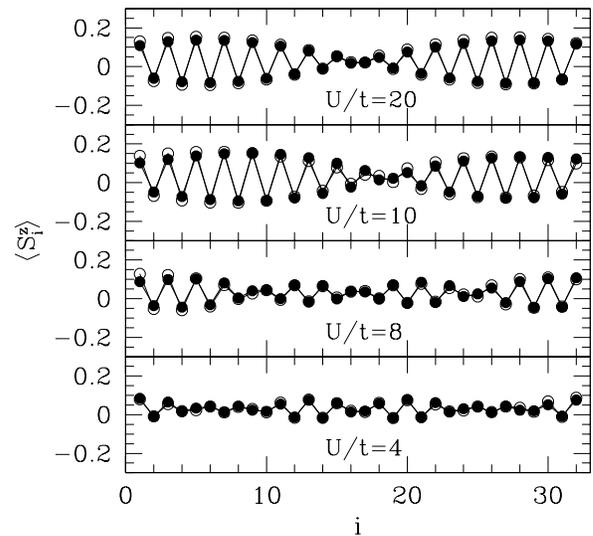}
\caption[DIAGONAL]{Spin distribution $\langle S^{z}_i\rangle$ in the $S^{z}_{\text{total}}=1$ state of the 
$t-t^{\prime}-U$ chain with $L=64$, $t^{\prime} = 3 t$ and
$U/t=4,8,10,20$. Black circles correspond to lattice odd and open circles to
even lattice sites.}
\label{Fig:2_Chain_SD}
\end{figure}

In Fig. \ref{Fig:2_Chain_SD}, we show the spin density distribution 
$\langle S^{z}_i\rangle$ in the $S^{z}_{\text{total}}=1$ state for
$t^{\prime} = 3 t$ and  $U/t=4,8,10,20$. 
As can be seen, the incommensurate
pattern of the spin distribution in the metallic phase at $U/t=4$ and $U/t=5$
transforms in the insulating phase ($U/t=10$ and $U/t=12$) into a
pattern which, for each chain, is similar to that of the single Heisenberg
chain in the commensurate phase.
This pattern, seen in, e.g., 
Fig.\ \ref{Fig:Multi_spinden_J1J2.eps} for $J^\prime/J=0,0.4,0.5$,
corresponds to two-spinon excitations, as previously discussed.
This indicates that the system behaves as two identical, weakly coupled 
$S=1/2$ Heisenberg chains at large $U$.

\section{Conclusions}

We have carried out a combined analytical and numerical analysis of the 
insulator-metal transition in the half-filled one-dimensional 
$t-t^{\prime}-U$ model. 
Using the weak-coupling bosonization approach, we have shown that the 
gross features of the transition from an insulator to metal as a
function of next-nearest-neighbor hopping
$t^{\prime}$ can be described within the standard theory of
commensurate-incommensurate transitions. 
We have derived an explicit expression for the critical line 
$t^{\prime}_{c}(U)$ separating the metallic phase from the spin-gapped
insulator.
We have also argued that in the vicinity of transition line a
crossover to infinite-order critical behavior occurs.

Using DMRG calculations on chains of up to $L=128$ sites, we have
performed a detailed numerical analysis of the excitation spectrum
and the charge and spin density distributions in various
sectors of the phase diagram. 
In particular, we have studied the evolution of the charge and spin gap with
increasing next-nearest-neighbor hopping amplitude $t^{\prime}$.
We have found evidence for a spin gap in the parameter range 
$0.5t<t^{\prime}<t^{\prime}_{c}$, in agreement with previous studies. 
We have shown that the change in the topology of Fermi surface at the
insulator-metal transition is reflected in the appearance of
incommensurate modulations of the charge density. 
Incommensurate spin-density distributions in the triplet sector are
always present in the metallic phase, but can also appear
independently in the spin-gapped insulator due to frustration.

For $t^{\prime} \gg t$, we have argued that the insulator-metal 
transition can be best understood starting from the limit of two
uncoupled chains.
At small $U$, turning on the zigzag coupling between the chains
destroys the commensurability present for a single chain, and leads to
a metallic phase.
At large $U$, the system is insulating and behaves as two weakly
coupled Heisenberg chains.
We have estimated that the insulator-metal transition 
in this regime occurs when the shift in the Fermi energy is comparable
to the size of the charge gap in the isolated Hubbard chain.

\acknowledgments{
It is a pleasure to thank A.\ Aligia, C.\ Gros, A.\ Kampf, B.\ Normand,
D.\ Poilblanc, and 
M.\ Sekania for interesting discussions.  
DB and GIJ also acknowledge support through the SCOPES grant 7GEPJ62379.
}


\begin{thebibliography}{99}
\bibitem{Mott_Book_90} N.\ F.\ Mott,
                 {\sl Metal-Insulator Transitions},
		 $2^{nd}$ ed., Taylor and Francis, London (1990).

\bibitem{Gebhard_Book_97} F.\ Gebhard,
                  {\sl The Mott Metal--Insulator Transition},
		  Springer, Berlin (1997).
		  
\bibitem{IFT_98} M.\ Imada, A.\ Fujimori, and Y.\ Tokura, 
                  Rev.\ Mod.\ Phys.\ {\bf 70}, 1039 (1998).

\bibitem{DBM_97}M.\ Dzierzawa, D.\ Baeriswyl and L.\ M.\ Martelo,
               Helv.\ Phys.\ Acta {\bf 70}, 124 (1997). 

\bibitem{LiebWu_68} E.\ H.\ Lieb and F.\ Y.\ Wu,
                 Phys.\ Rev.\ Lett.\ {\bf 20}, 1445 (1968).

\bibitem{AAO_1969} A.\ A.\ Ovchinnikov,
                 Sov.\ Phys.\ JETP {\bf 30}, 1160  (1970).

\bibitem{EFGKK_2005} For a recent review see F.\ H.\ L.\ Essler, H.\ Frahm, F.\ G\"ohmann,  
              A.~Kl\"umper, and V.\ E.\ Korepin, {\sl The One-Dimensional Hubbard Model}, 
              Cambridge Univ.\ Press, Cambridge (2005).

\bibitem{Fabrizio_96} M.\ Fabrizio,
                   Phys.\ Rev.\ B {\bf 54}, 10054 (1996).

\bibitem{Kuroki_97} K.\ Kuroki, R.\ Arita, and H.\ Aoki,
		J.\ Phys.\ Soc.\ Japan {\bf 66}, 3371 (1997).

\bibitem{DaulNoack_98} S.\ Daul and R.\ M.\ Noack,
                   Phys.\ Rev.\ B {\bf 58}, 2635 (1998).

\bibitem{Fabrizio_98} R.\ Arita, K.\ Kuroki, H.\ Aoki, and M.\ Fabrizio,
                   Phys.\ Rev.\ B {\bf 57}, 10324 (1998).

\bibitem{DaulNoack_00} S.\ Daul and R.\ M.\ Noack,
                   Phys.\ Rev.\ B {\bf 61}, 1646 (2000).

\bibitem{Torio_03} M.\ E.\ Torio, A.\ A.\ Aligia, and H.\ A.\ Ceccatto,
            Phys.\ Rev.\ B {\bf 67}, 165102 (2003).

\bibitem{AebBaerNoack_01} C.\ Aebischer, D.\ Baeriswyl, and R.\ M.\ Noack,
                Phys.\ Rev.\ Lett.\ {\bf 86}, 468 (2001).

\bibitem{Gros_01} K.\ Louis, J.\ V.\ Alvarez, and C.\ Gros,
                Phys.\ Rev.\ B {\bf 64}, 113106 (2001); {\bf 65}, 249903(E) (2002).

\bibitem{Gros_02} K.\ Hamacher, C.\ Gros, and W.\ Wenzel,
                Phys.\ Rev.\ Lett.\ {\bf 88}, 217203 (2002).

\bibitem{Gros_04} C.\ Gros, K.\ Hamacher, and W.\ Wenzel, 
                Europhys.\ Lett.\ {\bf 69}, 616 (2005).
	          
\bibitem{Fabrizio_04} M.\ Capello, F.\ Becca, M.\ Fabrizio, S.\ Sorella, and E.\ Tosatti,  
                       Phys.\ Rev.\ Lett.\ {\bf 94}, 026406 (2005).

\bibitem{C_IC_transition} G.\ I.\ Japaridze and A.\ A.\ Nersesyan, 
                  JETF Pis'ma {\bf 27}, 356 (1978); [JETP Lett.\ {\bf 27}, 334 (1978)]; 
                  J.\ Low Temp.\ Phys.\ {\bf 37}, 95 (1979);
                  V.\ L.\ Pokrovsky and A.\ L.\ Talapov,
                  Phys.\ Rev.\ Lett.\ {\bf 42}, 65 (1979);
                  H.\ J.\ Schulz, Phys.\ Rev.\ B {\bf 22}, 5274 (1980).

\bibitem{JNW_1984} G.\ I.\ Japaridze, A.\ A.\ Nersesyan, and P.\ B.\ Wiegmann,
                  Nucl.\ Phys.\ B {\bf 230}, 511 (1984).

\bibitem{Haldane_82} F.\ D.\ M.\ Haldane,
                   Phys.\ Rev.\ B {\bf 25}, 4925 (1982); {\bf 26}, 5257(E) (1982).

\bibitem{Eggert_96} K.\ Okamoto and K.\ Nomura,
		  Phys.\ Lett.\ A {\bf 169}, 422 (1992);
                    S.\ Eggert,
                  Phys.\ Rev.\ B {\bf 54}, R9612 (1996).

\bibitem{WhiteAffleck_96} S.\ R.\ White and I.\ Affleck,
                  Phys.\ Rev.\ B {\bf 54}, 9862 (1996).

\bibitem{BalentsFisher_96} L.\ Balents and M.\ P.\ A.\ Fisher,
                Phys.\ Rev.\ B {\bf 53}, 12133 (1996).

\bibitem{GNT} A.\ O.\ Gogolin, A.\ A.\ Nersesyan, and A.\ M.\ Tsvelik,
               {\it Bosonization and Strongly Correlated Systems},
               Cambridge Univ.\ Press, Cambridge (1998).

\bibitem{Giamarchi} T.\ Giamarchi, {\it Quantum Physics in One
               Dimension}, Clarendon Press, Oxford (2004).

\bibitem{note_1} For details we refer to  
                    Chapter 17 in Ref.\ \onlinecite{GNT}.

\bibitem{White_92} S.\ R.\ White, Phys.\ Rev.\ Lett.\ {\bf 69}, 2863 (1992);
                Phys.\ Rev.\ B {\bf 48}, 10345 (1993).

\bibitem{aebischer_thesis} C.\ Aebischer, Ph.\ D.\ thesis, University
                 of Fribourg, 2002, available at http://ethesis.unifr.ch/theses.

\bibitem{SAAP_1998} E.\ Sorensen, I.\ Affleck, D.\ Augier, and D.\ Poilblanc, 
                    Phys.\ Rev.\ B {\bf 58}, R14701 (1998).

\bibitem{Majumdar_Ghosh_1969}  C.\ K.\ Majumdar and D.\ K.\ Ghosh,
                               J.\ Math.\ Phys.\ {\bf 10}, 1388 (1969).


\end{thebibliography}
\end{document}